\documentclass[12pt,a4paper]{article}
\def\today{3 March 2000}
\usepackage{amsfonts}
\usepackage{latexsym}
\usepackage{geometry}

\usepackage{cite}
\def\bomega{{\mathord{\hbox{\boldmath$\omega$}}}}
\def\bOmega{{\mathord{\hbox{\boldmath$\Omega$}}}}

\def\bxi{{\mathord{\hbox{\boldmath$\xi$}}}}
\def\btheta{{\mathord{\hbox{\boldmath$\theta$}}}}

\def\SF{\mathcal{SF}}
\def\Vir{{\mathcal{V}ir}}
\def\i{\mathrm{i}}
\def\e{\mathrm{e}}
\def\d{\mathrm{d}}
\def\sign{\mathop\mathrm{sign}}
\def\tr{\mathop\mathrm{tr}\nolimits}
\begin{document}
%
%
\thispagestyle{empty}
\def\thefootnote{\fnsymbol{footnote}}
\begin{flushright}
  hep-th/0003029\\
  DTP/00/21
\end{flushright}
\vskip 2.5em
\begin{center}\LARGE
  Symplectic Fermions
\end{center}\vskip 2em
\begin{center}\large
  Horst G. Kausch%
  \footnote{Email: {\tt H.G.Kausch@dur.ac.uk}}%
\end{center}
\begin{center}\it
  Department of Mathematical Sciences, \\
  University of Durham, South Road, Durham DH1 3LE, U.K.
\end{center}
\vskip 1em
\begin{center}
  \today
\end{center}
\vskip 1em
\begin{abstract}
  We study two-dimensional conformal field theories generated from a
  ``symplectic fermion'' --- a free two-component fermion field of
  spin one --- and construct the maximal local supersymmetric
  conformal field theory generated from it.  This theory has central
  charge $c=-2$ and provides the simplest example of a theory with
  logarithmic operators.
  
  Twisted states with respect to the global
  $SL(2,\mathbb{C})$-symmetry of the symplectic fermions are
  introduced and we describe in detail how one obtains a consistent
  set of twisted amplitudes.  We study orbifold models with respect to
  finite subgroups of $SL(2,\mathbb{C})$ and obtain their modular
  invariant partition functions.  In the case of the cyclic orbifolds
  explicit expressions are given for all two-, three- and four-point
  functions of the fundamental fields.  The $\mathcal{C}_2$-orbifold
  is shown to be isomorphic to the bosonic local logarithmic conformal
  field theory of the triplet algebra encountered previously. We
  discuss the relation of the $\mathcal{C}_4$-orbifold to critical
  dense polymers.
  \begin{flushleft}
  \end{flushleft}
\end{abstract}

\setcounter{footnote}{0}
\def\thefootnote{\arabic{footnote}}


\section{Introduction}
\label{sec:intro}

Over the last few years it has become apparent that there is a class
of conformal field theories whose correlation functions have
logarithmic branch cuts. Such conformal field theories are believed to
be important for the description of certain statistical models, in
particular in the theory of (multi)critical polymers and percolation
\cite{Sal92,Flohr95,Kau95,MNR97,GLL99}, two-dimensional turbulence
\cite{RR95,Flohr96b,RR96}, and the quantum Hall effect
\cite{GurFloNay97,KT99}.
Models analysed so far include
the WZNW model on the supergroup $GL(1,1)$ \cite{RSal92}, the coset
model $SL(2,\mathbb{C})/SU(2)$ \cite{KT99}, the $c=-2$
model \cite{Gur93,GKau96a}, gravitationally dressed conformal field
theories \cite{BKog95} and some critical disordered models
\cite{CKT95,CKT96,MSer96,CTT98,GL99}. Singular vectors of some
Virasoro models have been constructed in \cite{Flohr98}, correlation
functions have been calculated in \cite{SR,RMK}, and more structural
properties of logarithmic conformal field theories have been analysed
in \cite{Roh}.

The class of chiral logarithmic conformal field theories is also
interesting from a conceptual point of view. There exist logarithmic
models which behave in many respects like ordinary (non-logarithmic)
chiral conformal field theories, and it is not yet clear in which way
these models differ structurally from conventional theories. They
include the models investigated in this paper, a series of
``quasirational logarithmic'' Virasoro models \cite{GKau96a} and a
series of ``rational logarithmic'' models, the simplest of which is
the triplet algebra at $c=-2$ \cite{GKau96b}. Here quasirational means
that a countable set of representations of the chiral algebra closes
under fusion (with finite fusion rules), and rational that the same
holds for a certain finite set of representations, including all
(finitely many) irreducible representations. For rational models Zhu's
algebra \cite{Zhu} is finite dimensional, and it should be possible to
read off all properties of the whole chiral theory from the vacuum
representation. In particular, one should be able to decide \textit{a
  priori} whether the chiral algebra leads to a logarithmic theory or
not, without actually constructing all the amplitudes. As yet little
progress has been made in this direction, although it is believed that
unitary (rational) meromorphic conformal field theories always lead to
non-logarithmic theories.

The only rational logarithmic model which has been studied in detail
so far, the aforementioned triplet algebra at $c=-2$, possesses
another oddity (apart from the appearance of indecomposable reducible
representations which lead to logarithmic correlation functions), and
it is quite possible that this is true in more generality
\cite{Flohr96a}: although the theory possesses a finite fusion
algebra, the matrices corresponding to the reducible representations
cannot be diagonalised, and a straightforward application of
Verlinde's formula does not make sense. This is mirrored by the fact
that the modular transformation properties of some of the characters
cannot be described by constant matrices as they depend on the modular
parameter $\tau$. This might suggest that these logarithmic rational
theories only make sense as chiral theories, and that they do not
correspond to modular invariant (non-chiral) conformal field theories.
It was demonstrated in \cite{GKau98} that, at least for the case of
the triplet algebra at $c=-2$, this is not the case. The resulting
theory is in every aspect a standard (non-chiral) conformal field
theory but for the property that it does not factorise into standard
chiral theories. Among other things, this demonstrates that a
non-chiral conformal field theory has significantly more structure
than the two chiral theories it is built from.

The general strategy for constructing a (non-chiral) conformal field
theory is as follows. Determine the two-, three- and four-point
functions of the \emph{fundamental} fields (the fields that correspond
to the fundamental states of the different representations). Given the
two- and three-point functions of the fundamental fields, all other
amplitudes can be derived from these, and the consistency conditions
of all amplitudes can be reduced to those being obeyed by the
four-point functions. This reduces the problem of constructing any
amplitude to a finite computation which can be done in principle. All
data of a conformal field theory can then be recovered from the
complete set of amplitudes, see \cite{GGod98}.  Due to the complex
structure of the non-chiral representations for the triplet model it
was not feasible in \cite{GKau98} to complete this programme. However,
the two- and three-point functions agree with those of a model built
from a two-component free fermion field, the ``symplectic fermion''
model.  Since the two- and three-point amplitudes uniquely determine
the theory, consistency of the symplectic fermion model implies
consistency of the triplet model.

Our strategy in this paper mirrors the one in \cite{GKau98}: Firstly
we construct the representations of the chiral algebra generated by
the symplectic fermions. In this case we obtain a unique maximal
indecomposable non-chiral representation.  The non-chiral amplitudes
are then determined as co-invariants with respect to the
comultiplication for the fermion field. After imposing the
consistency constraints on the hierarchy of amplitudes we arrive at a
unique local logarithmic conformal field theory
$\SF$ generated from the
symplectic fermions.  This theory admits a global $SL(2,\mathbb{C})$
symmetry under which the fermions transform in the fundamental
representation. We can thus introduce twisted states such that the
symplectic fermions acquire a phase when moved along a closed loop
around a twisted state. Amplitudes involving twisted states are still
co-invariants with respect to the (twisted) comultiplication for the
symplectic fermions and we shall use this to explicitly calculate the
local two-, three- and four-point amplitudes for twist fields. The
consistency constraints give a system of quadratic equations in the
three-point couplings which admits a simple solution.

Orbifold models $\SF[\mathcal{G}]$ with respect to a finite subgroup
of $SL(2,\mathbb{C})$ are obtained by restricting to (twisted and
untwisted) states which are invariant under $\mathcal{G}$. For the
cyclic subgroups this can be done explicitly with the local orbifold
amplitudes given by the above hierarchy of twisted amplitudes where
all fields are $\mathcal{G}$-invariant.  We determine the different
sectors of the models and give their modular invariant partition
functions.  The $\mathcal{C}_{2}$ orbifold model is the logarithmic
conformal field theory for the triplet algebra discussed in
\cite{GKau98} while the $\mathcal{C}_{4}$ orbifold model is related to
dense critical polymers \cite{Sal92}.  Our method cannot be applied to
the non-abelian orbifolds, for these we can only determine the
partition functions by path-integral arguments.

This paper is organised as follows. In Section~\ref{sec:symp} we
define the symplectic fermion model and construct its non-chiral
amplitudes. In Section~\ref{sec:twist} we introduce twisted sectors
and find their amplitudes. Finally, in Section~\ref{sec:orb} we
investigate the various orbifold models with respect to finite
subgroups of the global $SL(2,\mathbb{C})$ symmetry: the cyclic groups
in sections \ref{sec:c2n} and \ref{sec:cn} and non-abelian orbifolds
in section \ref{sec:naorb}.  The equivalence of the $\mathcal{C}_2$
orbifold with the triplet model is established in
Section~\ref{sec:trip} and the relation of the
$\mathcal{C}_4$-orbifold to critical dense polymers is discussed in
Section~\ref{sec:poly}.  The more technical aspects of the paper have
been relegated to a number of appendices. The locality of the
symplectic fermion model is established in Appendix~\ref{app:nonchir}.
The twisted amplitudes are explicitly constructed in
Appendix~\ref{app:tloc}.  The character of the orbifold chiral algebra
is derived in Appendix~\ref{app:chars}.

\section{Symplectic fermions}
\label{sec:symp}

We take as our starting point the $(\eta,\xi)$ ghost system with the first
order action \cite{FMSh86}
\begin{equation}
  \label{eq:etaxi}
  S = \frac{1}{\pi} \int \d^2z \left( 
    \eta\bar\partial\xi + \bar\eta\partial\bar\xi
  \right)
\end{equation}
where $\eta$ and $\xi$ are conjugate \emph{fermion} fields of dimensions
1 and 0, respectively.
The operator product of the chiral fields has short-distance limit
\begin{equation}
  \label{eq:ghostope}
  \eta(z)\xi(w) \sim \frac{1}{z-w}, \qquad
  \xi(z)\eta(w) \sim \frac{1}{z-w}
\end{equation}
up to terms regular as $z\to w$. 
The stress tensor for this system has central charge $c=-2$ and 
reads  
\begin{equation}
  \label{eq:stress}
  T(z) = 
  \mathopen: \partial\xi(z)\eta(z) \mathclose:,
\end{equation}
where $\mathopen:\cdots\mathclose:$ indicates fermionic normal
ordering.
The system also possesses a natural $U(1)$ current given by 
\begin{equation}
  \label{eq:current}
  J(z) = \mathopen: \xi(z)\eta(z) \mathclose:,
\end{equation}
which counts $\xi$ with charge $+1$ and $\eta$ with charge $-1$. 

Viewed as a conformal field theory by itself this system has a number
of problematic features.  If one wishes the vacuum state $\Omega$ to
be translation invariant the current is not a primary field
\cite{FMSh86}, rather one has
\begin{equation}
  \label{eq:current1}
  T(z) J(w) \sim \frac{-1}{(z-w)^3} + \frac{J(w)}{(z-w)^2} +
  \frac{\partial J(w)}{z-w}.
\end{equation}
As a consequence one has charge asymmetry,
$J_0^\dagger=1-J_0$. This implies anomalous charge conservation in
vacuum expectation values, 
\begin{equation}
  \label{eq:anomq}
  \langle\Omega| \psi_1(z_1,\bar z_1)\cdots\psi_n(z_n,\bar z_n)
  |\Omega\rangle = 0, 
  \qquad
  \textrm{unless $\Sigma q_i=\Sigma \bar{q}_i=1$},
\end{equation}
where $q_i$ and $\bar{q}_i$ are the charges of $\psi_i$ with respect
to the currents $J$ and $\bar{J}$. 
In terms of the fermions this means that vacuum expectation values are
non-vanishing only if they contain exactly one unpaired $\xi$ and
$\bar\xi$ field. One has two ground states, the invariant vacuum
$\Omega$ and $\xi_0\bar\xi_0\Omega$, with vanishing norms but
$\langle\Omega|\xi_0\bar\xi_0|\Omega\rangle=1$. Furthermore it is not
possible to construct an inner product on the space states compatible
with the standard hermiticity properties of the stress tensor,
$L_n^\dagger=L_{-n}$. 
A further problem is the appearance of fields which are neither
primary fields nor descendants of a primary field, the simplest
examples of which are $\xi$ and $J$; for further details see
\cite{Kau95}. 

The ghost system can be bosonised by defining $J(z)=\partial\phi(z)$
in which case the stress tensor takes on the Feigin-Fuchs form.
However, the aforementioned problems still persist. In the Coulomb gas
construction of the Virasoro minimal models they are resolved by
defining the physical states through a BRST resolution \cite{Feld89}.
This identifies the two ground states and removes the reducible
representations of the Virasoro algebra leaving only the field content
of the minimal model. The central charge $c=-2$ falls into the pattern
$c=1-6(p-q)^2/(pq)$ for the minimal series albeit with parameters
$p=1, q=2$ outside the allowed range. The screening charge for the
BRST resolution corresponds to $\eta_0$ and the usual ``physical''
space $\mathop\mathrm{ker}\eta_0/\mathop\mathrm{im}\eta_0$ is trivial,
as observed in \cite{Sal92}. This is reflected in the fact that the
standard Dotsenko-Fateev correlators \cite{DFat84} vanish identically
when evaluated for $c=-2$ due to cancellations between the conformal
blocks \cite{Sal92}. One can get finite results by considering the
limit $c\to-2$ however the the algebraic structure of the Virasoro
representations is discontinuous \cite{GKau96a}.

It was noted in \cite{FMSh86} that the $(\eta,\xi)$ system contains an
irreducible chiral algebra $\mathcal{A}$ generated by $\eta$ and
$\partial\xi$. Both of these are Virasoro primary fields of dimension
one.  We can put them on an equal footing by grouping them together as
a two-component fermion field $\chi^\alpha$ of dimension one.  We
shall now apply the algebraic methods introduced in \cite{GKau98} to
construct a local conformal field theory, the ``symplectic fermion
model'' $\SF$, containing the chiral algebra $\mathcal{A}$ generated
by $\chi^\alpha$.  The resulting model is generated by a
\emph{non-chiral} two-component free fermion field $\theta^\alpha$ of
dimension zero such that $\chi^\alpha=\partial\theta^\alpha,
\bar\chi^\alpha=\bar\partial\theta^\alpha$. In contrast, the
$(\eta,\xi)$ system can be obtained by adjoining to
$\mathcal{A}\otimes\bar\mathcal{A}$ the two \emph{chiral} fields $\xi$
and $\bar\xi$. The two models thus correspond to different local
slices of the same non-local theory.

\subsection{Chiral structure}
\label{sec:chir}

The chiral algebra $\mathcal{A}$ of the symplectic fermion model is
generated by a two-component fermion field,
\begin{equation}
  \chi^\alpha(z) = \sum_{n\in\mathbb{Z}} \chi^\alpha_n z^{-n-1} \,,
\end{equation}
of conformal weight one. 
The field has short-distance expansion 
\begin{equation}
  \label{eq:chiope}
  \chi^\alpha(z)\chi^\beta(w) \sim 
  \frac{d^{\alpha\beta}}{(z-w)^2}
\end{equation}
resulting in anti-commutators for the Fourier modes,
\begin{equation}
  \label{eq:chiac}
  \{ \chi^\alpha_m, \chi^\beta_n \} = m d^{\alpha\beta}
  \delta_{m+n}\,, 
\end{equation}
where $d^{\alpha\beta}$ is an anti-symmetric tensor; we may choose a
basis $\chi^\pm$ such that $d^{+-}=1$.  This algebra has a
\emph{unique} irreducible highest weight representation. Its highest
weight state $\Omega$ is annihilated by all non-negative fermion
modes, $\chi^\alpha_m \Omega = 0$ for $m\geq0$.  This representation
is isomorphic to $\mathcal{A}$ and provides its vacuum representation
with $\Omega$ as the vacuum state. The chiral algebra contains a
Virasoro algebra $\Vir$ of central charge $c=-2$ given by
\begin{equation}
  \label{eq:chiL}
  T(z) = 
  \sum_{n\in\mathbb{Z}} L_n z^{-n-2} = 
  \frac12 d_{\alpha\beta} 
  \mathopen: \chi^\alpha(z) \chi^\beta(z) \mathclose:,
\end{equation}
where $d_{\alpha\beta}$ is the inverse of $d^{\alpha\beta}$ such that
$d^{\alpha\beta}d_{\beta\gamma}=\delta^\alpha_\gamma$. In algebraic
language the stress tensor corresponds to the conformal state
\begin{equation}
  \label{eq:chiLs}
  L_{-2} \Omega = 
  \frac12 d_{\alpha\beta} \chi^\alpha_{-1} \chi^\beta_{-1} \Omega \,.
\end{equation}
The vacuum
$\Omega$ is M\"obius invariant under this Virasoro algebra,
$L_m\Omega=0$ for $m>-2$.

While the irreducible representation is unique it can be extended to
obtain reducible but indecomposable representations. Denote by
$\mathcal{A}^\sharp$ the maximal generalised highest weight
representation of $\mathcal{A}$ obtained as such an the extension of
the vacuum representation.  It is freely generated by the negative
modes, $\chi^\alpha_m$ for $m<0$, from a four dimensional space of
ground states.  This space is spanned by two bosonic states $\Omega$
and $\omega$, and two fermionic states, $\theta^\alpha$, with the
action of the zero-modes $\chi^\alpha_0$ given by
\begin{equation}
  \label{eq:chir0}
  \renewcommand{\arraystretch}{1.2}
  \begin{array}{rcl}
    \chi^\alpha_0 \omega &=& -\theta^\alpha\,, 
    \\
    \chi^\alpha_0 \theta^\beta &=& d^{\alpha\beta} \Omega\,, 
    \\
    L_0 \omega &=& \Omega\,. 
  \end{array}
\end{equation}
The states $\omega$ and $\Omega$ span a two-dimensional Jordan block
for $L_0$. A conformal field theory based on this maximal
representation will thus contain logarithmic operators. 

\subsection{Non-chiral theory}
\label{sec:nonchir}

To construct the non-chiral theory we proceed as described in
\cite{GKau98}. Details of the locality constraints are presented in
Appendix~\ref{app:nonchir}. The result is that there is a
\emph{unique} non-chiral symplectic fermion model $\SF$. It
has two bosonic ground states, $\Omega, \omega$, and two fermionic
ground states $\theta^\alpha$, satisfying
\begin{equation}
  \label{eq:nchir0}
  \renewcommand{\arraystretch}{1.2}
  \begin{array}{rcl}
    \chi^\alpha_0 \omega &=& 
    \bar\chi^\alpha_0 \omega = 
    -\theta^\alpha\,,
    \\
    \chi^\alpha_0 \theta^\beta &=& 
    \bar\chi^\alpha_0 \theta^\beta = 
    d^{\alpha\beta} \Omega\,,
    \\
    L_0 \omega &=& \bar L_0 \omega = \Omega\,.
  \end{array}
\end{equation}
The full non-chiral space of states $\mathcal{W}$ is generated by the
free action of the negative modes, $\chi^\alpha_m, \bar\chi^\alpha_m$
with $m<0$, from the ground states. Using the comultiplication
(\ref{eq:comult}), amplitudes of excited states can be expressed in
terms of amplitudes involving only fundamental fields, \textit{i.e.}
fields corresponding to the four ground states.  The amplitudes of up
to four fundamental fields are
\begin{eqnarray}
  \langle\omega\rangle &=& 1 \,, 
  \nonumber\\
  \langle\omega\omega\rangle &=& 
  -2 ({\circ}{-}{\circ}) \,, 
  \nonumber\\
  \langle\omega\omega\omega\rangle &=& 
  2 ({\circ}{-}{\circ}{-}{\circ}) - ({\circ}{=}{\circ}) \,, 
  \nonumber\\
  \langle\omega\omega\omega\omega\rangle &=& 
  2 ({\circ}{-}{\circ}{-}{\circ}{-}{\circ}) - 
  2 ({\circ}{=}{\circ}{\circ}{-}{\circ}) - 
  2 (\bigtriangledown) \,,
  \nonumber\\
  \langle\theta^\alpha\theta^\beta\rangle &=& d^{\alpha\beta} \,,
  \\
  \langle\theta^\alpha\theta^\beta\omega\rangle &=& 
  - d^{\alpha\beta} \left(
    \Delta_{13} + \Delta_{23} - \Delta_{12}
  \right)\,,
  \nonumber\\
  \langle\theta^\alpha\theta^\beta\omega\omega\rangle &=& 
  d^{\alpha\beta} \Bigl[
    (\Delta_{13}+\Delta_{14}+\Delta_{23}+\Delta_{24} -
    2\Delta_{12}-\Delta_{34}) \Delta_{34}
    \nonumber\\*&&\qquad{}
    - (\Delta_{13}-\Delta_{14})(\Delta_{23}-\Delta_{24})
  \Bigr],
  \nonumber\\
  \langle\theta^\alpha\theta^\beta\theta^\gamma\theta^\delta\rangle &=& 
  d^{\alpha\beta} d^{\gamma\delta} \left(\Delta_{12}+\Delta_{34}\right) 
  - 
  d^{\alpha\gamma} d^{\beta\delta} \left(\Delta_{13}+\Delta_{24}\right) 
  + 
  d^{\alpha\delta} d^{\beta\gamma} \left(\Delta_{14}+\Delta_{23}\right) 
  \,,
  \nonumber
\end{eqnarray}
where
\begin{displaymath}
  \begin{array}{rcl@{\qquad}rcl}
    ({\circ}{-}{\circ}) &=& \sum_{ij} \Delta_{ij}\,, 
    &
    ({\circ}{=}{\circ}) &=& \sum_{ij} \Delta_{ij}^2\,, 
    \\
    ({\circ}{-}{\circ}{-}{\circ}) &=& 
    \sum_{ijk} \Delta_{ij} \Delta_{jk}\,, 
    &
    ({\circ}{=}{\circ}{\circ}{-}{\circ}) &=& 
    \sum_{ijkl} \Delta_{ij}^2 \Delta_{kl}\,, 
    \\
    ({\circ}{-}{\circ}{-}{\circ}{-}{\circ}) &=& 
    \sum_{ijkl} \Delta_{ij} \Delta_{jk} \Delta_{kl}\,, 
    &
    (\bigtriangledown) &=& 
    \sum_{ijk} \Delta_{ij} \Delta_{jk} \Delta_{ki}\,
  \end{array}
\end{displaymath}
and 
\begin{equation}
  \label{eq:prop}
  \Delta_{ij} \equiv \Delta(z_{ij}) = 
  \mathcal{Z} + \ln|z_{ij}|^2 \,.
\end{equation}
The parameter $\mathcal{Z}$ could be set to zero by redefining
$\omega\mapsto\omega+\mathcal{Z}\Omega$, but it will be convenient
later on to retain this parameter.

The sums are over pairwise distinct labelled graphs ({\it i.e.} graphs
with vertices); labelled graphs that differ by a graph symmetry are
only counted once. In amplitudes, $\Omega$ acts as the unit operator,
except that its one-point function vanishes, $\langle\Omega\rangle=0$.
The relevant operator product expansions are, to lowest order in $x$, 
\begin{eqnarray}
  \theta^\alpha(x)\theta^\beta &\sim& 
  d^{\alpha\beta} \left(
    \omega + \Delta(x) \Omega
  \right) \,,
  \nonumber\\
  \theta^\alpha(x)\omega &\sim& 
  - \Delta(x) \theta^\alpha \,,
  \\
  \omega(x)\omega &\sim& 
  - \Delta(x) \left(
    2\omega + \Delta(x) \Omega
  \right) \,.
  \nonumber
\end{eqnarray}
The operator product of $\Omega$ with any field $S$ is simply given by
$\Omega(x)S = S$ to all orders.  

The structure of the theory is simple enough to allow us to determine
not just the four-point amplitudes but all $2N$-point amplitudes of
the $\theta$ fields.
\begin{equation}
  \label{eq:thetaamp}
  \langle\theta^{\alpha_1}\cdots\theta^{\alpha_{2N}}\rangle = 
  \frac{1}{2^{2N}}
  \sum_{\sigma\in\mathcal{S}_{2N}} \sign(\sigma)
  d^{\alpha_1\alpha_2} d^{\alpha_3\alpha_4} \cdots
  d^{\alpha_{2N-1}\alpha_{2N}} 
  \Delta_{\sigma_3\sigma_4}\cdots\Delta_{\sigma_{2N-1}\sigma_{2N}}
\end{equation}
where the sum is over all permutations of the fields and the prefactor
ensures that each grouping of the $2N$ fields into $N$ pairs is only
counted once. This can be proved by induction: the amplitudes satisfy
the correct differential equations in the $z_i$ and are thus fixed up
to addition of a constant; considering the OPE
$\theta^\alpha\theta^\beta$ fixes that constant.  By contracting two
$\theta$ fields one can obtain amplitudes involving $\omega$. In
particular, the $N$-point amplitude for $\omega$ can be written a sum
over graphs of $N-1$ links between the $N$ vertices,
\begin{equation}
  \label{eq:omegaamp}
  \langle\omega\cdots\omega\rangle = 
  \sum_{g\in\mathcal{G}^{N-1}_N} (-1)^{N+\mathcal{N}_c}
  2^{\mathcal{N}_d} 
  \Delta(g), 
\end{equation}
where the propagator $\Delta(g)$ for the graph $g$ is given as the
product of a propagator $\Delta_{ij}$ for each link between vertices
$i$ and $j$, $\mathcal{N}_c$ is the number of components including the
isolated vertex $\circ$ and the two-vertex loop ${\circ}{=}{\circ}$
while $\mathcal{N}_d$ is the number of components excluding $\circ$
and ${\circ}{=}{\circ}$. The combinatorial factor arises from
contracting the $d^{\alpha\beta}$ tensors in the $2N$-point amplitude
$\langle\theta\cdots\theta\rangle$ with the $d_{\alpha\beta}$ tensors
in the OPE of $\theta^\alpha\theta^\beta$ and factors as the product
of the following factors for each component of the graph $g$,
\begin{displaymath}
  \begin{array}{cc}
    {\circ} & 1 \\
    {\circ}{=}{\circ} & -1 \\
    {\circ}{-}{\circ}{-}\cdots{-}{\circ}{-}{\circ} & 2 (-1)^{n+1} \\
    \vtop{\ialign{#\crcr
        ${\circ}{-}{\circ}{-}\cdots{-}{\circ}{-}{\circ}$\crcr
        \noalign{\nointerlineskip}%
        $\kern3pt\vrule height5pt\hrulefill\vrule height5pt\kern3pt$\crcr}}
    & 2 (-1)^{n+1} 
  \end{array}
\end{displaymath}
Here the last two subgraphs contain $n$ vertices. For example, the
five- and six-point amplitudes of $\omega$ can be written as
\begin{eqnarray*}
  \langle\omega\omega\omega\omega\omega\rangle &=& 
  2({\circ}{-}{\circ}{-}{\circ}{-}{\circ}{-}{\circ})
  -2(\vtop{\ialign{#\crcr
      ${\circ}{-}{\circ}{-}{\circ}{-}{\circ}$\crcr
      \noalign{\nointerlineskip}%
      $\kern3pt\vrule height5pt\hrulefill\vrule
      height5pt\kern3pt$\crcr}})
  +({\circ}{=}{\circ}\,{\circ}{=}{\circ})
  \\*&&
  -4({\circ}{-}{\circ}\,\vtop{\ialign{#\crcr
      ${\circ}{-}{\circ}{-}{\circ}$\crcr
      \noalign{\nointerlineskip}%
      $\kern3pt\vrule height5pt\hrulefill\vrule
      height5pt\kern3pt$\crcr}})
  -2({\circ}{-}{\circ}{-}{\circ}\,{\circ}{=}{\circ})
  \\
  \langle\omega\omega\omega\omega\omega\omega\rangle &=& 
  -2({\circ}{-}{\circ}{-}{\circ}{-}{\circ}{-}{\circ}{-}{\circ})
  +2(\vtop{\ialign{#\crcr
      ${\circ}{-}{\circ}{-}{\circ}{-}{\circ}{-}{\circ}$\crcr
      \noalign{\nointerlineskip}%
      $\kern3pt\vrule height5pt\hrulefill\vrule
      height5pt\kern3pt$\crcr}})
  -2({\circ}{=}{\circ}\,\vtop{\ialign{#\crcr
      ${\circ}{-}{\circ}{-}{\circ}$\crcr
      \noalign{\nointerlineskip}%
      $\kern3pt\vrule height5pt\hrulefill\vrule
      height5pt\kern3pt$\crcr}})
  \\*&&
  +4({\circ}{-}{\circ}\,\vtop{\ialign{#\crcr
      ${\circ}{-}{\circ}{-}{\circ}{-}{\circ}$\crcr
      \noalign{\nointerlineskip}%
      $\kern3pt\vrule height5pt\hrulefill\vrule
      height5pt\kern3pt$\crcr}})
  +4({\circ}{-}{\circ}{-}{\circ}\,\vtop{\ialign{#\crcr
      ${\circ}{-}{\circ}{-}{\circ}$\crcr
      \noalign{\nointerlineskip}%
      $\kern3pt\vrule height5pt\hrulefill\vrule
      height5pt\kern3pt$\crcr}})
  -2({\circ}{-}{\circ}\,{\circ}{-}{\circ}\,{\circ}{=}{\circ})
  \\*&&
  +2({\circ}{-}{\circ}{-}{\circ}{-}{\circ}\,{\circ}{=}{\circ})
\end{eqnarray*}

These amplitudes define a fully consistent local conformal field
theory on the Riemann sphere.  This does not imply that the theory can
be defined consistently on higher genus Riemann surfaces. Indeed, the
partition function of this model is not modular invariant. To define
the theory on the torus one has to add additional sectors where the
fermion fields satisfy anti-periodic boundary conditions along the
fundamental cycles of the torus --- the different spin structures.
This will be discussed in more detail in Section \ref{sec:orb}.

\section{Twisted sectors}
\label{sec:twist}

The symplectic fermion model admits a global $SL(2,\mathbb{C})$
symmetry under which the fermion fields $\chi^\alpha$ transform in the
fundamental representation.  As in any theory where we have a global
group of automorphisms $\mathcal{G}$ we can introduce twist fields
$\tau_g$ for any $g\in\mathcal{G}$, such that, as a field $\phi$ from
$\mathcal{W}$ is taken around the insertion of a twist field $\tau_g$
we obtain the original field up to the action of the automorphism $g$,
\begin{equation}
  \phi(\e^{2\pi\i}z,\e^{-2\pi\i}\bar z)\tau_g =
  (g\phi)(z,\bar z) \tau_g \,.
\end{equation}
For the (chiral and anti-chiral) symplectic fermions this means,
\begin{equation}
  \chi^\alpha(\e^{2\pi\i}z)\tau_g =
  (g\chi^\alpha)(z) \tau_g \,,\qquad
  \bar\chi^\alpha(\e^{2\pi\i}\bar z)\tau_g =
  (g^{-1}\bar\chi^\alpha)(\bar z) \tau_g \,.
\end{equation}
We shall restrict to the case where $\mathcal{G}\subset U(1)$ is an
\emph{abelian} subgroup of $SL(2,\mathbb{C})$. In this case all the
twists commute and we can choose a basis $\chi^\pm$ for the two
fermion fields such that $d^{+-} = 1$ and the automorphisms are of the
form
\begin{equation}
  g\colon \chi^\pm \mapsto \e^{\pm2\pi\i\lambda} \chi^\pm
\end{equation}
for some $\lambda$ and we can use that parameter $\lambda$ to label
the different twisted sectors. It also follows that the symplectic
fermions have a mode expansion
\begin{equation}
  \label{eq:chitmode}
  \chi^\pm(z) = \sum_{n\in\mathbb{Z}} \chi^\pm_{-n\mp\lambda}
  z^{n-1\pm\lambda} \,,\qquad 
  \bar\chi^\pm(\bar z) = \sum_{n\in\mathbb{Z}}
  \bar\chi^\pm_{-n\pm\lambda} 
  \bar z^{n-1\mp\lambda} \,
\end{equation}
when acting on the $\lambda$-twisted sector.  Denote by
$\mathcal{W}_\lambda$ the representation freely generated from a
ground state $\mu_\lambda$ satisfying
$\chi^\alpha_r\mu_\lambda=\bar\chi^\alpha_r\mu_\lambda=0$ for $r>0$.
This twisted representation is irreducible since $\chi^\pm$ does not
have zero-modes.
The ground state $\mu_\lambda$ is a Virasoro highest weight state with
conformal weight
\begin{equation}
  \label{eq:twheight}
  h_\lambda = - \frac{\lambda(1-\lambda)}{2}.
\end{equation}
In this expression, and below, we always take $0<\lambda<1$.  We also
write $\lambda^*=1-\lambda$.  We will now construct the amplitudes
for the twisted sectors. We need to determine only the amplitudes of
the cyclic states $\omega$ and $\mu_\lambda$ since they determine all
other amplitudes through the twisted comultiplication
(\ref{eq:tcomult}). As usual, the system of differential equations
arising from the (twisted) comultiplication determines the amplitudes
up to some structure constants; the details can be found in
Appendix~\ref{app:tloc}.  Since the overall twist of an amplitude has
to vanish, all amplitudes with a single twist field vanish and the
non-vanishing amplitudes with two twist fields are
\begin{eqnarray}
  \label{eq:t2amp}
  \langle\mu_\lambda\mu_{\lambda^*}\rangle &=& 
  - \mathcal{O}_\lambda
  |z_{12}|^{2\lambda\lambda^*}, 
  \nonumber\\
  \langle\mu_\lambda\mu_{\lambda^*}\omega\rangle &=& 
  \phantom{-} \mathcal{O}_\lambda
  |z_{12}|^{2\lambda\lambda^*} \left(
    \mathcal{Z}_\lambda + \ln\left|\frac{z_{13}z_{23}}{z_{12}}\right|^2
  \right), 
  \\
  \langle\mu_\lambda\mu_{\lambda^*}\omega\omega\rangle &=&
  - \mathcal{O}_\lambda
  |z_{12}|^{2\lambda\lambda^*}
  \Biggl[
    \left(
      \mathcal{Z}_\lambda +
      \ln\left|\frac{z_{13}z_{23}}{z_{12}}\right|^2  
    \right) \left(
      \mathcal{Z}_\lambda +
      \ln\left|\frac{z_{14}z_{24}}{z_{12}}\right|^2  
    \right) 
    -
    H^\lambda(x)
    H^{\lambda^*}(x)
  \Biggr],
  \nonumber
\end{eqnarray}
where $H^\lambda(x)$ is given by
\begin{equation}
  \label{eq:Hlambda}
  H^\lambda(x,\bar x) = 
  - 2\Re \left( 
    \frac{(1-x)^{\lambda}}{\lambda}
    \,{}_2F_1(1,\lambda;1+\lambda;1-x)
  \right)
  + \mathcal{Y}_\lambda
  \,. 
\end{equation}
For each pair of twisted sectors we have two parameters,
$\mathcal{O}_\lambda$ and $\mathcal{Z}_\lambda$. 
The constant $\mathcal{Y}_\lambda$ is
fixed by locality of
$\langle\mu_\lambda\mu_{\lambda^*}\omega\omega\rangle$ and given
in Appendix~\ref{app:tloc2}. 
Amplitudes involving three twist fields are
\begin{equation}
  \label{eq:t3amp}
  \renewcommand{\arraystretch}{1.5}
  \begin{array}{rcll}
    \langle\mu_{\lambda_1}\mu_{\lambda_2}\mu_{\lambda_3}\rangle 
    &=&
    \mathcal{C}_{\lambda_1,\lambda_2,\lambda_3}
    \left|
      z_{12}^{\lambda_1\lambda_2}
      z_{13}^{\lambda_1\lambda_3}
      z_{23}^{\lambda_2\lambda_3}
    \right|^2
    &
    \hbox{for $\lambda_1+\lambda_2+\lambda_3=1$} \,, 
    \\
    &=&
    \mathcal{C}_{\lambda_1,\lambda_2,\lambda_3}
    \left|
      z_{12}^{\lambda_1^*\lambda_2^*}
      z_{13}^{\lambda_1^*\lambda_3^*}
      z_{23}^{\lambda_2^*\lambda_3^*}
    \right|^2
    &
    \hbox{for $\lambda_1+\lambda_2+\lambda_3=2$} \,, 
  \end{array}
\end{equation}
from which we obtain the OPEs
\begin{equation}
  \renewcommand{\arraystretch}{1.5}
  \begin{array}{rcll}
    \mu_{\lambda_1}(x)\mu_{\lambda_2}
    &=&
    - \frac{\mathcal{C}_{\lambda_1,\lambda_2,1-\lambda_1-\lambda_2}}
    {\mathcal{O}_{\lambda_1+\lambda_2}}
    |x|^{2\lambda_1\lambda_2}
    \mu_{\lambda_1+\lambda_2} + \cdots
    &
    \hbox{for $0<\lambda_1+\lambda_2<1$} \,, \\
    &=&
    - \frac{\mathcal{C}_{\lambda_1,\lambda_2,2-\lambda_1-\lambda_2}}
    {\mathcal{O}_{\lambda_1+\lambda_2-1}}
    |x|^{2\lambda_1^*\lambda_2^*}
    \mu_{\lambda_1+\lambda_2-1} + \cdots
    &
    \hbox{for $1<\lambda_1+\lambda_2<2$} \,, \\
    &=&
    - \frac{\mathcal{O}_{\lambda_1}}{\mathcal{O}}
    |x|^{2\lambda_1\lambda_2}
    \left(
      \omega + \Delta^{(\lambda)}(x) \Omega 
    \right) + \cdots
    &
    \hbox{for $\lambda_1+\lambda_2=1$} \,, 
  \end{array}
\end{equation}
where
\begin{equation}
  \label{eq:tprop}
  \Delta^{(\lambda)}(x) = 
  \ln|x|^2 + 2\mathcal{Z} - \mathcal{Z}_\lambda \,.
\end{equation}
Amplitudes involving four twist fields can be expressed in terms of
hypergeometric functions. As the expressions are quite complicated
they are listed in Appendix~\ref{app:tloc3}. 
By contracting two fields in these four-point amplitudes we can
express the four-point structure constants in terms of
the three-point couplings.
\begin{description}
\item[$\lambda_1+\lambda_2+\lambda_3+\lambda_4=1$:]
  \begin{displaymath}
    \mathcal{F}_{\lambda_1,\lambda_2,\lambda_3,\lambda_4}
    =
    \frac{
      \mathcal{C}_{\lambda_1,\lambda_2,1-\lambda_1-\lambda_2}
      \mathcal{C}_{\lambda_1+\lambda_2,\lambda_3,\lambda_4}
      }{\mathcal{O}_{\lambda_1+\lambda_2}}
  \end{displaymath}
\item[$\lambda_1+\lambda_2+\lambda_3+\lambda_4=3$:]
  \begin{displaymath}
    \mathcal{F}_{\lambda_1,\lambda_2,\lambda_3,\lambda_4}
    =
    \frac{
      \mathcal{C}_{\lambda_1,\lambda_2,2-\lambda_1-\lambda_2}
      \mathcal{C}_{\lambda_1+\lambda_2-1,\lambda_3,\lambda_4}
      }{\mathcal{O}_{\lambda_1+\lambda_2-1}}
  \end{displaymath}
\item[$\lambda_1+\lambda_2+\lambda_3+\lambda_4=2$:]
  \begin{displaymath}\renewcommand{\arraystretch}{2}
    \begin{array}{rcl@{\qquad}l}
      \mathcal{F}_{\lambda_1,\lambda_2,\lambda_3,\lambda_4}
      &=& \displaystyle
      \frac
      {\mathcal{C}_{\lambda_1,\lambda_2,1-\lambda_1-\lambda_2}
        \mathcal{C}_{\lambda_1+\lambda_2,\lambda_3,\lambda_4}}
      {\mathcal{O}_{\lambda_1+\lambda_2}
        \sqrt{\rho(\lambda_1,\lambda_2)\rho(\lambda_3^*,\lambda_4^*)}}
      & \hbox{if $\lambda_1+\lambda_2<1$}
      \\
      &=& \displaystyle
      \frac
      {\mathcal{C}_{\lambda_3,\lambda_4,1-\lambda_3-\lambda_4}
        \mathcal{C}_{\lambda_3+\lambda_4,\lambda_1,\lambda_2}}
      {\mathcal{O}_{\lambda_3+\lambda_4}
        \sqrt{\rho(\lambda_3,\lambda_4)\rho(\lambda_1^*,\lambda_2^*)}}
      & \hbox{if $\lambda_1+\lambda_2>1$}
      \\
      &=& \displaystyle
      \frac{\mathcal{O}_{\lambda_1}\mathcal{O}_{\lambda_3}}
      {\mathcal{O}} 
      & \hbox{if $\lambda_1+\lambda_2=1$}
    \end{array}
  \end{displaymath}
\end{description}
Here, $\rho(\lambda,\lambda')$ is a ratio of gamma functions,
\begin{equation}
  \label{eq:rho}
  \rho(\lambda,\lambda') =
  \frac{\Gamma(1-\lambda-\lambda')\Gamma(\lambda)\Gamma(\lambda')}
  {\Gamma(\lambda+\lambda')\Gamma(1-\lambda)\Gamma(1-\lambda')},
\end{equation}
and $\mathcal{F}_{\lambda_1,\lambda_2,\lambda_3,\lambda_4}$ is the
overall normalisation of the amplitude
$\langle\mu_{\lambda_1}\mu_{\lambda_2}\mu_{\lambda_3}\mu_{\lambda_3}\rangle$.
Locality also fixes the the parameter
\begin{equation}
  \label{eq:loccond}
  \mathcal{Z}_\lambda = 
  \mathcal{Z} + \vartheta_{\lambda,\lambda^*}, 
  \qquad
  \vartheta_{\lambda,\lambda'} =
  2\psi(1)-\psi(\lambda)-\psi(\lambda'), 
\end{equation}
where $\psi(x)=\Gamma(x)'/\Gamma(x)$ is the digamma function.  The
structure constants
$\mathcal{F}_{\lambda_1,\lambda_2,\lambda_3,\lambda_4}$ are completely
symmetric in their arguments and considering all possible orderings
yields quadratic constraints on the three-point couplings.  A solution
to these locality constraints is given by
\begin{equation}
  \label{eq:C}
  \mathcal{O}_\lambda = 1
  \,,\qquad
  \mathcal{C}_{\lambda_1,\lambda_2,\lambda_3} = 
  \mathcal{C}_{\lambda_1^*,\lambda_2^*,\lambda_3^*} = 
  \sqrt{\frac{\Gamma(\lambda_1)\Gamma(\lambda_2)
      \Gamma(\lambda_3)}
    {\Gamma(\lambda_1^*)\Gamma(\lambda_2^*)
      \Gamma(\lambda_3^*)}}
  \,, 
\end{equation}
where $\lambda_1+\lambda_2+\lambda_3=1$. The four-point couplings are 
\begin{equation}
  \label{eq:F}
  \mathcal{F}_{\lambda_1,\lambda_2,\lambda_3,\lambda_4} = 
  \mathcal{F}_{\lambda_1^*,\lambda_2^*,\lambda_3^*,\lambda_4^*} = 
  \sqrt{\frac{\Gamma(\lambda_1)\Gamma(\lambda_2)
      \Gamma(\lambda_3)\Gamma(\lambda_4)} 
    {\Gamma(\lambda_1^*)\Gamma(\lambda_2^*)
      \Gamma(\lambda_3^*)\Gamma(\lambda_4^*)}}
  \,,
\end{equation}
for $\lambda_1+\lambda_2+\lambda_3+\lambda_4=1$
and $\mathcal{F}_{\lambda_1,\lambda_2,\lambda_3,\lambda_4} = 1$ for
$\lambda_1+\lambda_2+\lambda_3+\lambda_4=2$.

This hierarchy of twisted amplitudes provides a consistent set of
semi-local amplitudes: By construction, the ground states of the
twisted sectors are $\mathcal{G}$-invariant and amplitudes involving
only those the ground states and the cyclic field $\omega$ are local
and satisfy all consistency constraints. However, excited states may
not be $\mathcal{G}$-invariant and their amplitudes will acquire phase
factors when moving fields along closed loops around such an excited
field. To obtain local amplitudes we have to restrict to
$\mathcal{G}$-invariant states; this defines the orbifold model
$\SF[\mathcal{G}]$.

\section{Orbifold models}
\label{sec:orb}

An abelian subgroup $\mathcal{G}$ of $SL(2,\mathbb{C}$ is either
$U(1)$ or $\mathcal{C}_N$, the cyclic group of order $N$. In all these
cases irreducible representations of $\mathcal{G}$ are one-dimensional
and can be labelled by a weight (or twist) $\lambda$ such that the
generator $g$ acts as $\exp(2\pi\i\lambda)$, as has been done
implicitly in the previous section. In the case of $\mathcal{C}_N$ the
twists are $\lambda\in\mathbb{Z}/N$ while for $U(1)$ the twists are
continuous, $\lambda\in\mathbb{R}$. In both cases the set of allowed
twists can be identified with $\mathcal{G}$ since shifting a twist by
an integer results in the same representation.

Given the symplectic fermion model $\SF$ with space of states
$\mathcal{W}$ and chiral algebra $\mathcal{A}$ we construct the
orbifold model $\SF[\mathcal{G}]$ as follows \cite{DVVV89}.  The space
of states is given by the $\mathcal{G}$-invariant subspaces
$\mathcal{H}_\lambda=\mathcal{W}_\lambda[\mathcal{G}]$ of the
(twisted) modules $\mathcal{W}_\lambda$ for all
$\lambda\in\mathcal{G}$. They form representations of the orbifold
chiral algebra $\mathcal{A}[\mathcal{G}]$. The orbifold amplitudes,
given by the above twisted amplitudes where all fields are
$\mathcal{G}$-invariant, are fully local.  The spaces
$\mathcal{H}_\lambda$ are in general reducible representations of the
orbifold chiral algebra $\mathcal{A}[\mathcal{G}]$ and decompose as
\begin{equation}
  \label{eq:Hdecomp}
  \mathcal{H}_\lambda = 
  \bigoplus_{\mu\in\mathcal{G}} \mathcal{H}_{\lambda,\mu}. 
\end{equation}
We first consider the $\Vir\times U(1)$ characters of the spaces
$\mathcal{W}$ and $\mathcal{W}_\lambda$. The characters of the
$\mathcal{G}$-invariant subspaces $\mathcal{H}_0$ and
$\mathcal{H}_{\lambda,\mu}$ can then be obtained by specialising the
$U(1)$ characters to the finite group $\mathcal{G}$ and using the
orthogonality of $\mathcal{G}$-characters.

Since $\mathcal{W}_\lambda$ is irreducible with respect to the
original chiral algebra $\mathcal{A}$ it is simply the product of
a chiral and an anti-chiral twisted sector, 
\begin{equation}
  \mathcal{W}_\lambda = 
  \mathcal{V}_\lambda \otimes \bar\mathcal{V}_{\lambda^*},
\end{equation}
where $\mathcal{V}_\lambda$ is generated from a ground state of
conformal weight $h_\lambda=-\lambda\lambda^*/2$ by the modes
$\chi^+_{-n-\lambda}$ and $\chi^-_{-n+\lambda-1}$ with $n\geq0$. 
These chiral spaces decompose under $\mathcal{G}$ as
\begin{equation}
  \mathcal{V}_\lambda = 
  \bigoplus_{\mu\in\mathcal{G}} \mathcal{V}_\lambda^\mu,
\end{equation}
such that $g$ acts as $\exp(2\pi\i\mu)$ on $\mathcal{V}_\lambda^\mu$.
The spaces $\mathcal{V}_\lambda^\mu$ are modules of the orbifold
chiral algebra $\mathcal{A}[\mathcal{G}]$ and we have chosen the
ground state to be invariant under $\mathcal{G}$.
We can introduce the chiral $\Vir\times U(1)$ character
\begin{eqnarray}
  \label{eq:vchar}
  \chi_{\mathcal{V}_\lambda}(\tau,g) &=& 
  \tr_{\mathcal{V}_\lambda}\left(
    \e^{2\pi\i\tau(L_0-c/24)} g
  \right)
  \nonumber\\
  &=&
  q^{-\frac{\lambda\lambda^*}{2}+\frac{1}{12}} \prod_{n=0}^\infty
  (1+gq^{n+\lambda}) (1+g^{-1}q^{n+\lambda^*})
  \\
  &=&
  \eta(\tau)^{-1} \sum_{m\in\mathbb{Z}} g^m q^{(2m+2\lambda-1)^2/8}
  \nonumber
\end{eqnarray}
where $\eta(\tau)$
is the Dedekind $\eta$-function,
$q=\exp(2\pi\i\tau)$ and $g$ is an element of $U(1)$. In the last line
we used the Jacobi triple product identity.
The untwisted sector $\mathcal{W}$ is not simply the product of a
chiral and an anti-chiral representation. However, since it is freely
generated from the ground states by the action of the chiral and
anti-chiral algebras, $\mathcal{A}$ and $\bar\mathcal{A}$, we have the
non-chiral $\Vir\times U(1)$ character
\begin{eqnarray}
  \label{eq:Wchar}
  \chi_{\mathcal{W}}(\tau,g) &=& 
  \tr_{\mathcal{V}_\lambda}\left(
    \e^{2\pi\i\tau(L_0-c/24)} \e^{-2\pi\i\bar\tau(\bar L_0-c/24)} g
  \right)
  \nonumber\\
  &=&
  (q\bar q)^{\frac{1}{12}} (2+g+g^{-1}) \left(
    \prod_{n=1}^\infty
    (1+gq^n) (1+g^{-1}q^n)
  \right) \left(
    \prod_{n=1}^\infty
    (1+g\bar q^n) (1+g^{-1}\bar q^n)
  \right) 
  \nonumber\\
  &=&
  \left( q^{\frac{1}{12}} 
    \prod_{n=0}^\infty
    (1+gq^{n}) (1+g^{-1}q^{n+1})
  \right) 
  \left( \bar q^{\frac{1}{12}} 
    \prod_{n=0}^\infty
    (1+g\bar q^{n+1}) (1+g^{-1}\bar q^{n})
  \right) 
  \nonumber\\
  &=&
  |\eta(\tau)|^{-2} 
  \left(
    \sum_{m\in\mathbb{Z}} g^m q^{(2m-1)^2/8}
  \right)
  \left(
    \sum_{m\in\mathbb{Z}} g^m \bar q^{(2m+1)^2/8}
  \right)
\end{eqnarray}

\subsection{$\mathcal{C}_{2N}$ orbifold}
\label{sec:c2n}

Specialising now to the case of $\mathcal{G}=\mathcal{C}_{2N}$ 
we have $g^{2N}=1$ for any group element $g$ and the twists are of the
form $\lambda=l/2N$. Thus the character of
$\mathcal{V}_{l/2N}$ can be written in the form
\begin{equation}
  \chi_{\mathcal{V}_{\frac{l}{2N}}}(\tau,g) = \sum_{k=0}^{2N-1} 
  g^k \Lambda_{2Nk-N+l, 2N^2}(\tau),
\end{equation}
where $\Lambda_{n,m}(\tau)=\Theta_{n,m}(\tau)/\eta(\tau)$ with the
classical theta function defined as
\begin{equation}
  \label{eq:theta}
  \Theta_{n,m}(\tau) = \sum_{k\in\mathbb{Z}+n/2m} q^{mk^2}.
\end{equation}
The orbifold representations $\mathcal{H}_\lambda =
\mathcal{W}_\lambda[\mathcal{C}_{2N}]$ with $\lambda\neq0$ are
reducible with respect to the orbifold chiral algebra
$\mathcal{A}[\mathcal{C}_{2N}]$,
\begin{equation}
  \label{eq:Hdecomp1}
  \mathcal{H}_\lambda = 
  \bigoplus_{\mu\in\mathcal{C}_{2N}} \mathcal{H}_{\lambda,\mu} = 
  \bigoplus_{\mu\in\mathcal{C}_{2N}} 
  \mathcal{V}_\lambda^\mu \otimes \bar\mathcal{V}_{\lambda^*}^{\mu^*}. 
\end{equation}
The non-chiral character of $\mathcal{H}_\lambda^\mu$ is just the
modulus squared of the chiral character of $\mathcal{V}_\lambda^\mu$,
\begin{equation}
  \label{eq:chiHlk}
  \chi_{\mathcal{H}_{l/2N,k/2N}}(\tau) = 
  \Bigl|
    \chi_{\mathcal{V}_{l/2N}^{k/2N}}(\tau) 
  \Bigr|^2 = 
  \left| \Lambda_{2Nk-N+l, 2N^2}(\tau) \right|^2.
\end{equation}
For the untwisted sector we obtain
\begin{eqnarray}
  \label{eq:chiH0}
  \chi_{\mathcal{H}_0}(\tau) &=& 
  |\eta(\tau)|^{-2} 
  \sum_{m-\bar m\equiv0(2N)} 
  q^{(2m-1)^2/8}\bar q^{(2\bar m-1)^2/8}
  \nonumber\\
  &=&
  |\eta(\tau)|^{-2} 
  \sum_{l=0}^{2N-1} 
  \left(
    \sum_{k\in\mathbb{Z}} q^{(4Nk+2l-1)^2/8}
  \right) \left(
    \sum_{\bar k\in\mathbb{Z}} \bar q^{(4N\bar k+2l-1)^2/8}
  \right)
  \nonumber\\
  &=&
  \sum_{l=0}^{2N-1} |\Lambda_{(2l-1)N,2N^2}(\tau)|^2
\end{eqnarray}
The space of states of the orbifold theory
$\SF[\mathcal{C}_{2N}]$ consists of the
indecomposable (extended) vacuum module $\mathcal{H}_0$, arising from
the untwisted sector $\mathcal{W}$, 
and $2N(2N-1)$ sectors 
\begin{equation}
  \mathcal{H}_{\lambda,\mu} = 
  \mathcal{V}_\lambda^\mu \otimes \bar\mathcal{V}_{\lambda^*}^{\mu^*},
\end{equation}
with $\lambda\neq0$, arising from the decomposition 
\begin{equation}
  \mathcal{H}_\lambda = 
  \mathcal{W}_\lambda[\mathcal{C}_{2N}] = 
  \bigoplus_{\mu\in\mathcal{C}_{2N}} \mathcal{H}_{\lambda,\mu}. 
\end{equation}
of the twisted sectors $\mathcal{W}_\lambda$.
The partition function of
$\SF[\mathcal{C}_{2N}]$ is thus
\begin{eqnarray}
  \label{eq:ZC2N}
  Z[\mathcal{C}_{2N}] &=& 
  \chi_{\mathcal{H}_0}(\tau) + 
  \sum_{l=1}^{2N-1} \sum_{k=0}^{2N-1}
  \chi_{\mathcal{H}_{l/2N,k/2N}}(\tau) 
  \nonumber\\*
  &=& 
  \sum_{l=0}^{2N-1} \sum_{k=0}^{2N-1}
  \left| \Lambda_{2Nk-N+l, 2N^2}(\tau) \right|^2
  \nonumber\\*
  &=& 
  \sum_{m=0}^{4N^2-1} 
  \left| \Lambda_{m, 2N^2}(\tau) \right|^2.
\end{eqnarray}
This partition function is invariant under the modular group and is in
fact identical to the partition function of a free boson compactified
on a circle of radius $r=2N$ using the normalisation of \cite{Gin88}.
Furthermore, defining as in \cite{IDro89} the Coulomb gas partition
function at radius $\rho$ as
\begin{equation}
  \label{eq:coul}
  Z(\rho,\tau) = \frac{1}{|\eta(\tau)|^2} \sum_{m,n\in\mathbb{Z}}
  q^{\frac{1}{4}(m\rho+n/\rho)^2} 
  \bar q^{\frac{1}{4}(m\rho-n/\rho)^2} 
\end{equation}
we have $Z[\mathcal{C}_{2N}](\tau)=Z(\sqrt{2}N,\tau)$. 

The orbifold chiral algebra $\mathcal{A}[\mathcal{C}_{2N}]$ has
character (see Appendix~\ref{app:chars})
\begin{equation}
  \label{eq:chiA}
  \chi_{\mathcal{A}[\mathcal{C}_{2N}]}(\tau) 
  =
  \frac{1}{2N} \Biggl[
    \eta(\tau)^2 + 
    \sum_{l=0}^{N-1} (-1)^{l} (2N-2l-1)
    \Lambda_{(2l+1)N,2N^2}(\tau) 
  \Biggr]
\end{equation}
Because of $\eta(-1/\tau)^2 = -\tau\eta(\tau)^2$, this does not
transform nicely under the modular group. This is an indication that
the chiral algebra $\mathcal{A}[\mathcal{C}_{2N}]$ should not by
itself form a sector of the theory but rather be contained within the
larger vacuum module $\mathcal{H}_0$. 

We can also decompose the chiral algebra $\mathcal{A}$ with respect to
the Virasoro algebra of central charge $c=-2$ contained within it. The
character of $\mathcal{A}$ can then be expressed in terms of the irreducible
characters of the Virasoro algebra at $c=-2$ as
\begin{equation}
  \chi_{\mathcal{A}}(\tau,u) 
  =
  \sum_{n=0}^\infty \left( \sum_{l=0}^n u^{2l-n} \right)
  \chi^{\Vir}_{n+1,1}(\tau),
\end{equation}
and thus the character of the orbifold chiral algebra is given in
terms of Virasoro characters as
\begin{equation}
  \chi_{\mathcal{A}[\mathcal{C}_{2N}]}(\tau,u) 
  =
  \sum_{k=0}^\infty 
  \left( 1 + 2 \left\lfloor \frac{k}{N} \right\rfloor \right)
  \chi^{\Vir}_{2k+1,1}(\tau),
\end{equation}
where $\lfloor x \rfloor$ is the largest integer less or equal to
$x$. Viewed as a $W$-algebra, the orbifold chiral algebra
$\mathcal{A}[\mathcal{C}_{2N}]$ is generated by the Virasoro field $L$
of conformal weight two, a Virasoro primary field of weight three and
a pair of Virasoro primary fields of weight $N(2N+1)$. In the case of
$N=1$ the three Virasoro primary fields all have weight three and we
obtain the triplet algebra, see Section \ref{sec:trip}.

\subsection{$\mathcal{C}_{N}$ orbifold}
\label{sec:cn}

We can repeat the same procedure for $\mathcal{C}_{N}$ orbifolds with
$N$ odd. However, the $\mathcal{C}_{N}$-invariant states
still contain fermions. The partition function $Z[\mathcal{C}_N]$
corresponds to a single spin-structure and is not modular invariant.
The $\mathcal{C}_{N}$ orbifold has an additional $\mathcal{C}_2$
symmetry arising from the boson-fermion selection rule. Performing the
sum over spin-structures in the $\SF[\mathcal{C}_{N}]$ model is the
same as directly performing the orbifold construction with respect to
$\mathcal{C}_{N}\times\mathcal{C}_{2}=\mathcal{C}_{2N}$.

\subsection{Non-abelian Orbifolds}
\label{sec:naorb}

The remaining finite subgroups of $SL(2,\mathbb{C})$ are the binary
tetrahedral, octahedral and icosahedral groups
$\mathcal{T},\mathcal{O}$ and $\mathcal{I}$ and one would like to
construct $\SF[\mathcal{G}]$ orbifolds for them as well. The general
structure of orbifold models was investigated in \cite{DVVV89}: The
twisted sectors are labelled by conjugacy classes of $\mathcal{G}$.
They can be constructed as before and, in fact, the structure of the
$g$-twisted module depends only on the order of $g$. The $n$-point
amplitudes are labelled by $n$-tuples of group elements
$(g_1,\ldots,g_n)$ such that $g_1\cdots g_n=1$. Labels related by
simultaneous conjugation yield the same amplitude.  Our method for
constructing the amplitudes relies on being able to simultaneously
diagonalise the twist for all fields in the amplitude. This is not
possible if some of the $g_i$ are not mutually commuting as is
necessarily the case for some amplitudes when $\mathcal{G}$ is
non-abelian. A direct construction of the $\mathcal{T},\mathcal{O}$
and $\mathcal{I}$ orbifolds is thus not possible.

However, we can determine the torus partition function of
$\SF[\mathcal{G}]$ as a sum over partition functions of the symplectic
fermion field $\theta^\alpha$ with different boundary conditions: For
$g,h\in\mathcal{G}$ we denote the partition function of the $h$-twisted
sector with an insertion of the operator $g$ as
\begin{equation}
  g \mathop{\Box}\limits_{\textstyle h} =
  \tr_{\mathcal{W}_h}\left( g
  q^{L_0-\frac{c}{24}} \bar q^{\bar L_0-\frac{c}{24}} \right) .
\end{equation}
The $\mathcal{G}$-orbifold partition function is then obtained as
\begin{equation}
  Z[\mathcal{G}] = \frac1{|\mathcal{G}|} 
  \sum_{\textstyle{g,h\in\mathcal{G}\atop gh=hg}} 
  g \mathop{\Box}\limits_{\textstyle h}.
\end{equation}
For non-abelian $\mathcal{G}$ boundary conditions twisted by
non-commuting group elements are not consistent, hence the condition
$gh=hg$.
This allows us to interpret the twisted partition functions as the
$\Vir\times U(1)$ characters introduced previously,
\begin{equation}
  \label{eq:tZ}
  g \mathop{\Box}\limits_{\textstyle h} = \chi_{\mathcal{W}_h}(\tau,g).
\end{equation}
For $\mathcal{C}_{2N}$ we recover the previous result which we denote
by $Z_N = Z[\mathcal{C}_{2N}]$. 
To calculate the partition function $Z[\mathcal{G}]$ for non-abelian
$\mathcal{G}$ we follow \cite{Gins88} and add the contributions of the
mutually commuting subsets of $\mathcal{G}$, which form cyclic groups,
subtracting any overcounting. As in \cite{Gins88} the result is
\begin{eqnarray}
 \label{eq:ZDN}
  Z[\mathcal{D}_N] &=& \frac12(Z_N + 2Z_2 - Z_1), \\
 \label{eq:ZT}
  Z[\mathcal{T}] &=& \frac12(2Z_3 + Z_2 - Z_1), \\
 \label{eq:ZO}
  Z[\mathcal{O}] &=& \frac12(Z_4 + Z_3 + Z_2 - Z_1), \\
 \label{eq:ZI}
  Z[\mathcal{I}] &=& \frac12(Z_5 + Z_3 + Z_2 - Z_1). 
\end{eqnarray}
Here, of course, one has to keep in mind that $Z_N(\tau)$ corresponds
to a Coulomb gas partition function at radius $N\sqrt2$ and not $N$ as
for $c=1$. 

\subsection{The triplet model}
\label{sec:trip}

We will now show that the $\mathcal{C}_2$-twisted model is the
logarithmic conformal field theory of \cite{GKau98}.  The chiral
algebra of the theory in \cite{GKau98} is a $W$-algebra, the so-called
triplet algebra \cite{Kausch91,Kau95}; it is realised in the
symplectic fermion model by
\begin{equation}
  \renewcommand{\arraystretch}{1.5}
  \label{eq:trip}
  \begin{array}{rcl}
    L_{-2} \Omega &=& 
    \displaystyle
    \frac12 d_{\alpha\beta} \chi^\alpha_{-1} \chi^\beta_{-1} \Omega \,, 
    \\
    W^a_{-3} \Omega &=& 
    t^a_{\alpha\beta} \chi^\alpha_{-2} \chi^\beta_{-1} \Omega \,.
  \end{array}
\end{equation}
Here the matrices $(t^a)^\alpha_\beta$ form the spin $1/2$
representation of $sl(2)$ in a Cartan-Weyl basis,
\begin{equation}
  t^{0\pm}_\pm = \pm\frac{1}{2}, \qquad
  t^{\pm\mp}_\pm=1,
\end{equation}
with all other entries vanishing. 
The bosonic sector $\mathcal{H}_0=\mathcal{W}_0[\mathcal{C}_2]$ of
$\mathcal{W}_0$ contains the representation $\mathcal{R}$ of the
triplet algebra \cite{GKau98}.  Explicitly, the ground states $\Omega$
and $\omega$ are identified in both representations and the higher
level states of $\mathcal{R}$ can be expressed as fermionic
descendents as
\begin{equation}
  \label{eq:Rdesc}
  \renewcommand{\arraystretch}{1.2}
  \begin{array}{rcl}
    \rho^{\alpha\bar\alpha} &=& \chi^\alpha_{-1}
    \bar\theta^{\bar\alpha}\,, \\
    \bar\rho^{\alpha\bar\alpha} &=& -\bar\chi^{\bar\alpha}_{-1}
    \theta^\alpha\,, \\
    \psi^{\alpha\bar\alpha} &=& 
    \chi^\alpha_{-1}\bar\chi^{\bar\alpha}_{-1}\Omega\,, \\
    \phi^{\alpha\bar\alpha} &=& 
    \chi^\alpha_{-1}\bar\chi^{\bar\alpha}_{-1}\omega\,.
  \end{array}
\end{equation}
On the other hand, the bosonic sector has the non-chiral character
\begin{equation}
  \chi_{\mathcal{H}_0}(\tau) 
  =
  2 \left| \Lambda_{1,2}(\tau) \right|^2 
  =
  \chi_{\mathcal{R}}(\tau) \,.
\end{equation}
Since $\mathcal{H}_0$ and $\mathcal{R}$ have the same character but
$\mathcal{H}_0$ contains $\mathcal{R}$ they are in fact identical,
$\mathcal{H}_0=\mathcal{R}$. By the same character argument we find
that the orbifold algebra $\mathcal{A}[\mathcal{C}_2]$ is identical to
the triplet algebra. 

The other two representations of the triplet model, the irreducible
representations $\mathcal{ V}_{-1/8,-1/8}$ and
$\mathcal{V}_{3/8,3/8}$, also have an interpretation in terms of the
symplectic fermion theory: they correspond to the bosonic sector
$\mathcal{H}_{\frac12}=\mathcal{W}_{\frac12}[\mathcal{C}_2]$ of the
(unique) $\mathcal{C}_2$-twisted representation
$\mathcal{W}_{\frac12}$. In this sector, the fermions are
half-integrally moded, but all bosonic operators (including the
triplet algebra generators that are bilinear in the fermions) are
still integrally moded, and the twisted sector decomposes as
\begin{equation}
  \mathcal{H}_{\frac{1}{2}} = 
  \mathcal{H}_{\frac{1}{2},0} \oplus
  \mathcal{H}_{\frac{1}{2},\frac{1}{2}}. 
\end{equation}
with respect to the triplet algebra.  The ground state $\mu$ of the
twisted sector is identified with the ground state of $\mathcal{
  V}_{-1/8,-1/8}$ while the ground state of $\mathcal{V}_{3/8,3/8}$ is
given as a fermionic descendant in $\mathcal{H}_{\frac12}$,
\begin{equation}
  \nu^{\alpha\bar\alpha} = 
  \chi^\alpha_{-\frac12}\bar\chi^{\bar\alpha}_{\frac12}\mu\,. 
\end{equation}
Thus, $\mathcal{ V}_{-1/8,-1/8}$ is contained in
$\mathcal{H}_{\frac12,0}$ and $\mathcal{V}_{3/8,3/8}$ is contained in
$\mathcal{H}_{\frac12,\frac12}$ and by considering the characters,
\begin{equation}
  \renewcommand{\arraystretch}{1.2}
  \begin{array}{rcl}
    \chi_{\mathcal{H}_{\frac12,0}}(\tau) 
    &=& 
    |\Lambda_{0,2}(\tau)|^2 
    = \chi_{\mathcal{V}_{-1/8,-1/8}}(\tau)\,, 
    \\
    \chi_{\mathcal{H}_{\frac12,\frac12}}(\tau) 
    &=& 
    |\Lambda_{2,2}(\tau)|^2 
    =
    \chi_{\mathcal{V}_{3/8,3/8}}(\tau)\,,
  \end{array}
\end{equation}
we find that, in fact,
$\mathcal{H}_{\frac12,0}=\mathcal{V}_{-1/8,-1/8}$ and
$\mathcal{H}_{\frac12,\frac12}=\mathcal{V}_{3/8,3/8}$ as
representations of the triplet algebra. The $\mathcal{C}_2$ orbifold
and the triplet theory have the same partition function
\begin{equation}
  \label{eq:ZC2}
  Z[\mathcal{C}_2] = 
  \sum_{k=0}^3 \left|\Lambda_{k,2}(\tau)\right|^2 
  =
  Z_{\mathrm{triplet}} \,.
\end{equation}

The amplitudes determined in Section \ref{sec:twist} agree with those
in \cite{GKau98} on setting $\mathcal{Z}=4\ln2$ and thus
$\mathcal{Z}_{1/2}=8\ln2$ . Furthermore, all three-point functions of
the fundamental fields, $\rho^{\alpha\bar\alpha},
\bar\rho^{\alpha\bar\alpha}, \psi^{\alpha\bar\alpha}$ and
$\phi^{\alpha\bar\alpha}$, of the triplet model agree with the
corresponding excited amplitudes in the $\mathcal{C}_2$-twisted
symplectic fermion model. Both models are therefore isomorphic.  This
is the argument used in \cite{GKau98} to establish the consistency of
the logarithmic theory presented there.

\subsection{Critical dense polymers}
\label{sec:poly}

Many properties of polymers in solution can be modeled by considering
simple geometrical systems. In particular, dense polymers are obtained
by considering a finite number of self-avoiding and mutually-avoiding
loops or chains on a lattice that cover a finite fraction of the
available volume.  It was argued in \cite{Sal92} that their continuum
limit should correspond to a $(\xi,\eta)$ system. Specifically, the
$(\xi,\eta)$ system on a torus with periodic or anti-periodic boundary
conditions describes the sector formed by an even number of
non-contractible loops while the sector formed by an odd number of
non-contractible loops corresponds to $\mathbb{Z}_4$-twisted boundary
conditions. The total polymer partition function is equal to the
partition function $Z[\mathcal{C}_4]$.  This result was obtained by
realising dense polymers as the $n\to0$ limit of the low temperature
phase of the $O(n)$ model which in turn can be mapped onto a Coulomb
gas.  This also reproduces the scaling dimensions
\begin{equation}
  \label{eq:scale}
  x_L^D = \frac{L^2-4}{16}
\end{equation}
for the geometric polymer $L$-leg operators $\Phi_L$. 
However,
the limit $n\to0$ does not commute with the thermodynamic limit or
with changing the boundary conditions. 
Furthermore, the physical quantities which have been determined for
dense polymers, the partition function $Z[\mathcal{C}_4]$ and the
scaling dimensions (\ref{eq:scale}), are shared by the $\mathcal{C}_4$
orbifold models of both the $(\xi,\eta)$ system and the symplectic
fermions. Both models necessarily involve reducible but indecomposable
representations of the Virasoro algebra. 
The structure of these representations is quite different in the two
models resulting in differences for the correlators. In particular,
some amplitudes such as the four-point amplitude of the Ramond ground
states of dimensions $-1/8$ vanish when calculated directly in the
$(\xi,\eta)$ system or in its Coulomb gas formulation. To get a
non-zero result one has to take the generic Coulomb gas amplitude and
take the limit $c\to-2$ resulting in an amplitude with logarithmic
short-distance behaviour. This agrees with the amplitude
$\langle\mu_{1/2}\mu_{1/2}\mu_{1/2}\mu_{1/2}\rangle$ for the
symplectic fermions. The same is also the case for the other
logarithmic amplitudes thus providing evidence that the correct
description of the continuum theory of dense polymers is in terms of
the $\SF[\mathcal{C}_4]$ model. However, a more detailed analysis is
clearly needed: The 1-leg operator $\Phi_1$ is two-fold degenerate and
is represented by $\mu_{1/4}$ and $\mu_{3/4}$. The reason for this
degeneracy given in \cite{Sal92} is that sources and sinks of polymers
are distinguished.  The four-point amplitude of the 1-leg operator
$\Phi_1$ (eq. (88) of \cite{Sal92}) agrees with the amplitude
$\langle\mu_{1/4}\mu_{3/4}\mu_{1/4}\mu_{3/4}\rangle$ of the symplectic
fermion model. However, we also obtain a non-vanishing amplitude for
four sources (or four sinks),
\begin{displaymath}
  \langle\mu_{1/4}\mu_{1/4}\mu_{1/4}\mu_{1/4}\rangle = 
  \langle\mu_{3/4}\mu_{3/4}\mu_{3/4}\mu_{3/4}\rangle = 
  \frac{\Gamma(1/4)^2}{\Gamma(3/4)^2}
  |z_{12}z_{13}z_{14}z_{23}z_{24}z_{34}|^{1/8}.
\end{displaymath}
An important problem is to understand the role played by the
logarithmic fields and identify their equivalents in the lattice
realisation.  In both the $\mathbb{Z}_4$-twisted $(\xi,\eta)$-system
and the orbifold model $\SF[\mathcal{C}_4]$ there are two states of
conformal dimension $h=\bar h=0$. According to \cite{Sal92} these
correspond to the identity and the density operator $\rho$. In the
$(\xi,\eta)$-system the density operator is represented by
$\rho=\xi+\bar\xi$ and results in non-vanishing expectation value
$\langle\rho\rangle$ and zero density-density correlation
$\langle\rho\rho\rangle=0$. In the symplectic fermion model the
density would be some linear-combination of $\Omega$ and $\omega$
resulting again in non-vanishing expectation value but the the
density-density correlation acquires a logarithmic behaviour

\begin{displaymath}
  \langle\rho(z_1)\rho(z_2)\rangle = A + B \ln|z_{12}|^2.
\end{displaymath}
For a detailed comparison with lattice correlators it would be useful
to determine the amplitudes of the $\SF[\mathcal{C}_4]$ in finite
geometries, \textit{e.g.} a rectangular region or strip with various
boundary conditions. The short-distance behaviour of these amplitudes
will agree with those on the Riemann sphere calculated here but the
global behaviour will be modified by finite-size corrections.

The symplectic fermion model provides definite predictions for polymer
correlators going beyond just the set of scaling dimensions.  It is
hoped that numerical lattice simulations will be able to verify these
predictions and distinguish them from the $\mathbb{Z}_4$-twisted
$(\xi,\eta)$ system. On the lattice the logarithmic behaviour of the
correlators may however be masked by discretisation effects.

Similar considerations apply to dilute polymer or percolation models
whose continuum limit is described a $c=0$ conformal field theory. The
scaling dimensions for dilute polymers can be reproduced by the Kac
formula of conformal weights for $c=0$ if one also allows half-integer
indices \cite{Sal92}. A conformal field theory with central charge
$c=0$ containing Virasoro primary of these conformal weights cannot be
the Virasoro minimal model at $c=0$ since this has empty field
content. By an analysis of fusion products analogous to that described
in \cite{GKau96a} one can show that a $c=0$ model which contains fields
from the Kac table has to include logarithmic fields.
Preliminary results indicate that the generalised highest weight
representations occuring have a considerably more complicated
structure than in the $c=-2$ case. Further work in this direction will
be reported elsewhere. 

\paragraph{Acknowledgements:}

The author would like to thank Matthias Gaberdiel for many useful
discussions.


\appendix
\section{Locality of symplectic fermions}
\label{app:nonchir}

To construct non-chiral local representations we proceed as described
in \cite{GKau98}. We start off with the tensor product
$\mathcal{A}^\sharp\otimes\bar\mathcal{A}^\sharp$ of the
left and right chiral representations. In a local theory the operator
$S=L_0^{(\mathrm{n})}-\bar L_0^{(\mathrm{n})}$, \textit{i.e.} the
nilpotent part of $L_0-\bar L_0$, has to vanish on all states.  The
(maximal) non-chiral representation $\mathcal{W}_{\mathrm{max}}$ is
thus given as the quotient space
\begin{equation}
  \label{eq:Wmax}
  \mathcal{W}_{\mathrm{max}} =
  \left(
    \mathcal{A}^\sharp\otimes\bar\mathcal{A}^\sharp
  \right) / \mathcal{N}, 
\end{equation}
where $\mathcal{N}$ is the subrepresentation generated from
$S(\omega\otimes\bar\omega)$. The space of ground states then has the
structure
\begin{equation}
  \label{eq:nchir0b}
  \begin{array}{rcl@{\qquad}rcl}
    \chi^\alpha_0 \bomega &=& -\btheta^\alpha\,, &
    \bar\chi^{\bar\alpha}_0 \bomega &=& -\bar\btheta^{\bar\alpha}\,, 
    \\
    \chi^\alpha_0 \btheta^\beta &=& d^{\alpha\beta} \bOmega\,, &
    \bar\chi^{\bar\alpha}_0 \bar\btheta^{\bar\beta} &=& 
    d^{\alpha\beta} \bOmega\,, 
    \\
    \chi^\alpha_0 \bar\btheta^{\bar\alpha} &=&
    -\bxi^{\alpha\bar\alpha} \,, &
    \bar\chi^{\bar\alpha}_0 \btheta^\alpha &=&
    \bxi^{\alpha\bar\alpha} \,.
  \end{array}
\end{equation}
Here, $\bomega$ is the equivalence class of states in
$\mathcal{W}_{\mathrm{max}}$ which contains $\omega\otimes\bar\omega$
as a representative.  Since $S$ commutes with both chiral algebras and
$\mathcal{A}^\sharp$ is freely generated from the ground space
representation by the negative modes $\chi^\alpha_m,
\bar\chi^{\bar\alpha}_m$ with $m<0$, the same is true of
$\mathcal{W}_{\mathrm{max}}$.

The representation $\mathcal{W}_{\mathrm{max}}$ contains the states
which are allowed \textit{a priori} in a local theory.  However,
the space of states $\mathcal{W}$ actually realised in the non-chiral
local theory may be smaller than $\mathcal{W}_{\mathrm{max}}$, as
we might be forced to set some of those states to zero when
requiring the locality of two- and three-point amplitudes. 
We construct these amplitudes by the method described in
\cite{GKau98}: The $N$-point amplitudes are co-invariants with respect
to the comultiplications $\Delta^i(\phi_{-n})$, where $\phi$ is any
field in the chiral algebra, $n>-h$ with $h$ the conformal weight
of $\phi$ and $i=1,\ldots,N$. In our case it is sufficient to use the
comultiplications
\begin{equation}
  \label{eq:comult}
  \Delta^i(\chi^\alpha_{-n}) = \chi^{\alpha(i)}_{-n} + 
  \sum_{j\neq i} \varepsilon_{ij} \sum_{k=0}^\infty {-n\choose k}
  z_{ji}^{-n-k} \chi^{\alpha(j)}_k,
\end{equation}
where $\chi^{\alpha(j)}_k$ is the mode $\chi^\alpha_k$ acting on the
$j$-th field in the amplitude and $\varepsilon_{ij}$ is a sign factor
arising from interchanging of fermions when moving $\chi^\alpha$ from
$i$-th to $j$-th position. The comultiplication allows us to express
any amplitude in terms of amplitudes of the ground states alone. These
in turn satisfy systems of first order differential equations obtained
by identifying $\d/\d{z_i}$ with $L_{-1}$ acting on the $i$-th field
and noting that
\begin{equation}
  \label{eq:Lm1}
  L_{-1} \bOmega = 0, \qquad 
  L_{-1} \bomega = -d_{\alpha\beta} \chi^\alpha_{-1}\btheta^\beta,
  \qquad 
  L_{-1} \btheta^\alpha = \chi^\alpha_{-1}\bOmega.
\end{equation}
In this way the bosonic two-point amplitudes are found as
\begin{eqnarray}
  \langle\bomega\bomega\rangle &=& 
  -2\mathcal{C}_1 - 2 \mathcal{C}_0 \ln|z_{12}|^2, 
  \nonumber\\
  \langle\bomega\bOmega\rangle &=& \mathcal{C}_0, 
  \nonumber\\
  \langle\bomega\bxi^{\alpha\bar\alpha}\rangle &=& 
  -\mathcal{C}_0 \Theta^{\alpha\bar\alpha},
  \\
  \langle\bOmega\bOmega\rangle &=& 0, 
  \nonumber\\
  \langle\bOmega\bxi^{\alpha\bar\alpha}\rangle &=& 0, 
  \nonumber\\
  \langle\bxi^{\alpha\bar\alpha}\bxi^{\beta\bar\beta}\rangle &=& 0,
  \nonumber
\end{eqnarray}
while the amplitudes of two fermionic fields are
\begin{eqnarray}
  \langle\btheta^\alpha\btheta^\beta\rangle &=&
  d^{\alpha\beta} \mathcal{C}_0, 
  \nonumber\\
  \langle\bar\btheta^{\bar\alpha}\bar\btheta^{\bar\beta}\rangle &=&
  d^{\bar\alpha\bar\beta} \mathcal{C}_0, 
  \\
  \langle\btheta^\alpha\bar\btheta^{\bar\alpha}\rangle &=&
  -\Theta^{\alpha\bar\alpha} \mathcal{C}_0 . 
  \nonumber
\end{eqnarray}
Here, $\mathcal{C}_0, \mathcal{C}_1$ and $\Theta^{\alpha\bar\alpha}$
are arbitrary constants. Amplitudes of one fermionic and one bosonic
field vanish.  These amplitudes imply that $\bxi^{\alpha\bar\alpha}$
and $\bOmega$ are linearly dependent and thus\footnote{We set states
  to zero whose amplitudes vanish identically.}
\begin{equation}
  \bxi^{\alpha\bar\alpha} = - \Theta^{\alpha\bar\alpha} \bOmega.
\end{equation}
The fermionic states $\btheta^\alpha$ and $\bar\btheta^{\bar\alpha}$
are linearly dependent provided $\det(\Theta)=1$. This is in fact
required by locality of the amplitude
$\langle\btheta^\alpha\btheta^\beta\bomega\rangle$:
Solving the system of differential equations arising from the
comultiplication we obtain
\begin{equation}
  \langle\btheta^\alpha\btheta^\beta\bomega\rangle = 
  -d^{\alpha\beta} \left(
    \mathcal{A}_1 + 
    \mathcal{A}_0 \ln\frac{z_{13}z_{23}}{z_{12}} + 
    \det(\Theta) 
    \mathcal{A}_0 \ln\frac{\bar z_{13}\bar z_{23}}{\bar z_{12}}
  \right) . 
\end{equation}
This amplitude can be local only if $\det(\Theta)=1$.
Then the
four fermionic states $\btheta^\alpha$ and $\bar\btheta^{\bar\alpha}$
are related as
\begin{equation}
  \btheta^\alpha = 
  \Theta^{\alpha\bar\alpha} d_{\bar\alpha\bar\beta}
  \bar\btheta^{\bar\beta}\,, \qquad
  \bar\btheta^{\bar\alpha} = 
  - \Theta^{\alpha\bar\alpha} d_{\alpha\beta} \btheta^\beta\,. 
\end{equation}
The space of states $\mathcal{W}$ is then freely generated by the
negative modes, $\chi^\alpha_m$ and $\bar\chi^\alpha_m$ with $m<0$,
from the four ground states; two bosonic
states, $\bOmega$ and $\bomega$, and two fermionic states $\btheta^\alpha$

We now determine the remaining three-point amplitudes.
\begin{eqnarray}
  \langle\bOmega\bOmega\bOmega\rangle &=& 0, 
  \nonumber\\
  \langle\bomega\bOmega\bOmega\rangle &=& \mathcal{A}_0, 
  \nonumber\\
  \langle\bomega\bomega\bOmega\rangle &=& 
  -2\mathcal{A}_1 - 2 \mathcal{A}_0 \ln|z_{12}|^2, 
  \nonumber\\
  \langle\bomega\bomega\bomega\rangle &=& 
  3\mathcal{A}_2 + 2\mathcal{A}_1 \ln|z_{12}z_{13}z_{23}|^2 
  \\*&&+ 
  2 \mathcal{A}_0 \left(
    \ln|z_{12}|^2\ln|z_{13}|^2 + 
    \ln|z_{12}|^2\ln|z_{23}|^2 + 
    \ln|z_{13}|^2\ln|z_{23}|^2
  \right) 
  \nonumber\\*&&- 
  \mathcal{A}_0 \left(
    (\ln|z_{12}|^2)^2 + 
    (\ln|z_{13}|^2)^2 + 
    (\ln|z_{23}|^2)^2
  \right), 
  \nonumber\\
  \langle\btheta^\alpha\btheta^\beta\bOmega\rangle &=& 
  d^{\alpha\beta} \mathcal{A}_0, 
  \nonumber\\
  \langle\btheta^\alpha\btheta^\beta\bomega\rangle &=& 
  -d^{\alpha\beta} \left(
    \mathcal{A}_1 + \mathcal{A}_0 \ln\left|\frac{z_{13}z_{23}}{z_{12}}\right|^2
  \right). 
  \nonumber
\end{eqnarray}
From the two- and three-point amplitudes we can read off the
OPEs. Locality requires that the different ways of contracting two
fields in a three-point amplitude are equivalent; this implies
\begin{equation}
  \frac{\mathcal{C}_1}{\mathcal{C}_0} = 
  \frac{\mathcal{A}_1}{\mathcal{A}_0} = 
  \frac{\mathcal{A}_2}{\mathcal{A}_1} = 
  \mathcal{Z}.
\end{equation}
We further introduce parameters $\Lambda$ and $\mathcal{O}$ by
$\mathcal{C}_0=\Lambda^4\mathcal{O},
\mathcal{A}_0=\Lambda^6\mathcal{O}$. 
The OPEs are then
\begin{eqnarray}
  \Lambda^{-2} 
  \btheta^\alpha(x)\btheta^\beta &=& 
  d^{\alpha\beta} \left(
    \bomega + (\mathcal{Z} + \ln|x|^2) \bOmega
  \right), 
  \nonumber\\
  \Lambda^{-2} 
  \btheta^\alpha(x)\bomega &=& 
  - (\mathcal{Z} + \ln|x|^2) \btheta^\alpha, 
  \\
  \Lambda^{-2} 
  \bomega(x)\bomega &=& 
  - (\mathcal{Z} + \ln|x|^2) \left(
    2\bomega + (\mathcal{Z} + \ln|x|^2) \bOmega
  \right), 
  \nonumber
\end{eqnarray}
The operator product of $\bOmega$ with any field $\mathbf{S}$ is
simply given by $\bOmega(x)\mathbf{S} = \Lambda^2\mathbf{S}$ to all
orders. For $\Lambda^2=1$, the fields $\bOmega$ can be thought as the
unit operator, except that its one-point function vanishes,
$\langle\bOmega\rangle=0$. 

The two- and three-point amplitudes completely determine the theory
and all higher amplitudes can be constructed, at least in principle,
from these by a gluing process. Conversely, by taking limits of
$n$-point amplitudes in which two fields are close together, we can
relate $n$-point amplitudes to $(n-1)$-point amplitudes and the OPEs.
The different ways of so contracting two fields all have to be
compatible and this leads to consistency relations. It is sufficient
to check these for the four-point amplitudes. These are listed in the
main part of the paper and do indeed satisfy all consistency
relations. This shows that the symplectic fermions define a local
conformal field theory.

In the main part of this paper we revert to regular greek letters
instead of bold ones to denote the non-chiral fields. We also fix the
parameters as $\Lambda=\Omega=1$. They could easily be restored by
multiplying all $n$-point amplitudes by $\Lambda^{2n}\mathcal{O}$.
Furthermore, by performing a global chiral (or anti-chiral) $SU(2)$
transformation, we can choose $\Theta^{\alpha\bar\alpha} =
d^{\alpha\bar\alpha}$. The only remaining parameter, $\mathcal{Z}$,
corresponds to the freedom of adding to $\bomega$ an arbitrary
multiple of $\bOmega$. 

\section{Twisted amplitudes}
\label{app:tloc}

In abelian orbifolds the twist conditions satisfied by the twist
fields in any amplitude all commute. This implies that the twists can
be simultaneously diagonalised and insertion points of twist fields are
rational branch points for the symplectic fermions. To calculate
amplitudes involving twisted fields we can then use a twisted
comultiplication for the chiral fermion fields. To derive the
comultiplication formula consider (non-chiral) fields $\phi_j$ having
twists $\alpha_j$ with respect to a chiral field $S(w)$. If
$\alpha_1+\alpha_2+\alpha_3+\alpha_4\in\mathbb{Z}$ then, as a function
of $w$,
\begin{equation}
  (w-z_1)^{-\alpha_1} (w-z_2)^{-\alpha_2} (w-z_3)^{-\alpha_3}
  (w-z_4)^{-\alpha_4} \langle S(w)\phi_1\phi_2\phi_3\phi_4\rangle , 
\end{equation}
is meromorphic with poles at $w=z_j$. Taking a contour integral of $w$
around the other fields we obtain the comultiplication formula for
$S$, 
\begin{eqnarray}
  \label{eq:tcomult}
  \lefteqn{
  \Delta(S_{-\alpha_1-\alpha_2-\alpha_3-\alpha_4-h_S+1}) = 
  } \qquad && \nonumber\\*&&
  \sum_{r,s,t=0}^\infty 
  {-\alpha_2\choose r} {-\alpha_3\choose s} {-\alpha_4\choose t} 
  z_{12}^{-\alpha_2-r} z_{13}^{-\alpha_3-s} z_{14}^{-\alpha_4-t}
  S^{(1)}_{-\alpha_1+r+s+t-h_S+1}
  \nonumber\\*&&+
  \sum_{r,s,t=0}^\infty 
  {-\alpha_1\choose r} {-\alpha_3\choose s} {-\alpha_4\choose t} 
  z_{21}^{-\alpha_1-r} z_{23}^{-\alpha_3-s} z_{24}^{-\alpha_4-t}
  S^{(2)}_{-\alpha_2+r+s+t-h_S+1}
  \\*&&+
  \sum_{r,s,t=0}^\infty 
  {-\alpha_1\choose r} {-\alpha_2\choose s} {-\alpha_4\choose t} 
  z_{31}^{-\alpha_1-r} z_{32}^{-\alpha_2-s} z_{34}^{-\alpha_4-t}
  S^{(3)}_{-\alpha_3+r+s+t-h_S+1}
  \nonumber\\*&&+
  \sum_{r,s,t=0}^\infty 
  {-\alpha_1\choose r} {-\alpha_2\choose s} {-\alpha_3\choose t} 
  z_{41}^{-\alpha_1-r} z_{42}^{-\alpha_2-s} z_{43}^{-\alpha_3-t}
  S^{(4)}_{-\alpha_4+r+s+t-h_S+1}
  ,
  \nonumber
\end{eqnarray}
where $S^{(j)}_m$ acts on the $j$th field. If some fields are
fermionic there are additional minus signs from the interchange of two
fermion fields. As in the untwisted case any amplitude functional
$\Phi$ satisfies
\begin{equation}
  \Phi\circ\Delta(S_m) = 0 
  \qquad\hbox{for $m<h_S$,}
\end{equation}
that is, $\sum \alpha_j \geq-2(h_S-1)$. 
Two- and three-point amplitudes satisfy analogous comultiplication
properties. 

We can use the comultiplication to derive differential equations for
the amplitudes. Note that
\begin{equation}
  \renewcommand{\arraystretch}{1.2}
  \begin{array}{rcl}
    L_{-1} \mu &=& \chi^-_{\lambda-1} \chi^+_{-\lambda} \mu \,,
    \\
    L_{-1}^2 \mu &=& 
    (1-\lambda) \chi^-_{\lambda-2} \chi^+_{-\lambda} \mu
    -
    \lambda \chi^+_{-1-\lambda} \chi^-_{\lambda-1} \mu \,.
  \end{array}
\end{equation}
To derive a differential equation for
$\langle\mu_{\lambda_1}\cdots\rangle$ we typically use the
comultiplication for $\chi^-$ with $\alpha_1=1-\lambda_1$ and all
other $\alpha_j\leq0$ such that $\sum_j \alpha_j=0$ to change the 
$\chi^-_{\lambda-1}$ mode on the first field into $\chi^-_0$ acting on
the fields with $\alpha_j=0$. A second application of the
comultiplication for $\chi^+$ with $\alpha_1=\lambda_1$ and all
other $\alpha_j\leq0$ such that $\sum_j \alpha_j=0$ then changes the 
$\chi^+_{-\lambda}$ mode on the first field into $\chi^-_+$ acting on
the fields with $\alpha_j=0$. In cases where we cannot satisfy these
conditions on the $\alpha_j$ we can derive a second order differential
equation instead, as detailed below.

\subsection{Two twist fields}
\label{app:tloc2}

Amplitudes with two twist fields,
$\langle\mu_{\lambda_1}\mu_{\lambda_2}\cdots\rangle$ are non-zero only
if $\lambda_1+\lambda_2=1$. They satisfy first order differential
equations, for example,
\begin{eqnarray}
  \left( \partial_1 - \frac{\lambda\lambda^*}{z_{12}} \right)
  \langle\mu_{\lambda}\mu_{\lambda^*}\rangle &=& 0
  \,,
  \\
  \left( \partial_1 - \frac{\lambda\lambda^*}{z_{12}} \right)
  \langle\mu_{\lambda}\mu_{\lambda^*}\omega\rangle &=&
  \frac{z_{23}}{z_{12}z_{13}}
  \langle\mu_{\lambda}\mu_{\lambda^*}\Omega\rangle 
  \,,
  \\
  \left( \partial_1 - \frac{\lambda\lambda^*}{z_{12}} \right)
  \langle\mu_{\lambda}\mu_{\lambda^*}\theta^+\theta^-\rangle &=&
  \frac{z_{23}}{z_{12}z_{13}}
  \left(\frac{z_{13}z_{24}}{z_{14}z_{23}}\right)^{\lambda}
  \langle\mu_{\lambda}\mu_{\lambda^*}\Omega\Omega\rangle 
  \,,
  \\
  \left( \partial_1 - \frac{\lambda\lambda^*}{z_{12}} \right)
  \langle\mu_{\lambda}\mu_{\lambda^*}\omega\omega\rangle &=&
  \frac{z_{23}}{z_{12}z_{13}}
  \langle\mu_{\lambda}\mu_{\lambda^*}\Omega\omega\rangle 
  +
  \frac{z_{24}}{z_{12}z_{14}}
  \langle\mu_{\lambda}\mu_{\lambda^*}\omega\Omega\rangle 
  \nonumber\\*&&
  +
  \frac{z_{23}}{z_{12}z_{13}}
  \left(\frac{z_{13}z_{24}}{z_{14}z_{23}}\right)^{\lambda}
  \langle\mu_{\lambda}\mu_{\lambda^*}\theta^-\theta^+\rangle 
  \\*&&
  -
  \frac{z_{24}}{z_{12}z_{14}}
  \left(\frac{z_{14}z_{23}}{z_{13}z_{24}}\right)^{\lambda}
  \langle\mu_{\lambda}\mu_{\lambda^*}\theta^+\theta^-\rangle 
  \,.
  \nonumber
\end{eqnarray}
Imposing M\"obius covariance and monodromy invariance fixes the
functional form of the amplitudes,
\begin{eqnarray}
  \langle\mu_{\lambda}\mu_{\lambda^*}\rangle &=& 
  \mathcal{D}_\lambda |z_{12}|^{2\lambda\lambda^*}, 
  \\
  \langle\mu_{\lambda}\mu_{\lambda^*}\Omega\rangle &=& 
  \mathcal{D}^{(1)}_\lambda |z_{12}|^{2\lambda\lambda^*}, 
  \\
  \langle\mu_{\lambda}\mu_{\lambda^*}\omega\rangle &=& 
  - \mathcal{D}^{(1)}_\lambda |z_{12}|^{2\lambda\lambda^*}
  \left(
    \mathcal{Z}_\lambda +
    \ln\left|\frac{z_{13}z_{23}}{z_{12}}\right|^2 
  \right), 
  \\
  \langle\mu_{\lambda}\mu_{\lambda^*}\Omega\Omega\rangle &=& 
  \mathcal{D}^{(2)}_\lambda |z_{12}|^{2\lambda\lambda^*}, 
  \\
  \langle\mu_{\lambda}\mu_{\lambda^*}\omega\Omega\rangle &=& 
  - \mathcal{D}^{(2)}_\lambda |z_{12}|^{2\lambda\lambda^*}
  \left(
    \mathcal{Z}^{(1)}_\lambda +
    \ln\left|\frac{z_{13}z_{23}}{z_{12}}\right|^2 
  \right), 
  \\
  \langle\mu_{\lambda}\mu_{\lambda^*}\Omega\omega\rangle &=& 
  - \mathcal{D}^{(2)}_\lambda |z_{12}|^{2\lambda\lambda^*}
  \left(
    \mathcal{Z}^{(2)}_\lambda +
    \ln\left|\frac{z_{14}z_{24}}{z_{12}}\right|^2 
  \right), 
  \\
  \langle\mu_{\lambda}\mu_{\lambda^*}\theta^+\theta^-\rangle &=& 
  \mathcal{D}^{(2)}_\lambda |z_{12}|^{2\lambda\lambda^*}
  H^\lambda(x,\bar x), 
  \\
  \langle\mu_{\lambda}\mu_{\lambda^*}\theta^-\theta^+\rangle &=& 
  - \mathcal{D}^{(2)}_\lambda |z_{12}|^{2\lambda\lambda^*}
  H^{\lambda^*}(x,\bar x), 
  \\
  \langle\mu_{\lambda}\mu_{\lambda^*}\omega\omega\rangle &=& 
  \mathcal{D}^{(2)}_\lambda |z_{12}|^{2\lambda\lambda^*}
  \Biggl[
    \Biggl(
      \mathcal{Z}^{(1)}_\lambda +
      \ln\left|\frac{z_{13}z_{23}}{z_{12}}\right|^2 
    \Biggr)  
    \Biggl(
      \mathcal{Z}^{(2)}_\lambda +
      \ln\left|\frac{z_{14}z_{24}}{z_{12}}\right|^2 
    \Biggr) 
    \nonumber\\*&&\quad{}
    -
    H^\lambda(x,\bar x) 
    H^{\lambda^*}(x,\bar x)
    + \mathcal{X}_\lambda
  \Biggr], 
\end{eqnarray}
where $H^\lambda(x,\bar x)$ is given by
\begin{eqnarray}
  H^\lambda(x,\bar x) &=& 
  - 2 \Re\left(
    \frac{(1-x)^{\lambda}}{\lambda}
    \,{}_2F_1(1,\lambda;1+\lambda;1-x)
  \right)
  + \mathcal{Y}_\lambda 
  \\*
  &=&
  \ln|x|^2 - 2\vartheta_{1,\lambda} + \mathcal{Y}_\lambda 
  + 
  2 \Re\left(
    (1-x)^{\lambda} M(1,\lambda;1;x)
  \right)
  \nonumber
\end{eqnarray}
Reading off the OPEs from the two- and three-point amplitudes and
imposing locality on the four-point amplitudes fixes the constants
appearing in the amplitudes in terms of two new free parameters,
$\mathcal{O}_\lambda$ and $\mathcal{Z}_\lambda$, for each pair of
conjugate twisted sectors, 
\begin{equation}
  \renewcommand{\arraystretch}{1.5}
  \begin{array}{c}
    \mathcal{D}_\lambda = - \Lambda^2 \mathcal{O}_\lambda,
    \qquad
    \mathcal{D}^{(1)}_\lambda = - \Lambda^4 \mathcal{O}_\lambda,
    \qquad
    \mathcal{D}^{(2)}_\lambda = - \Lambda^6 \mathcal{O}_\lambda,
    \\
    \mathcal{Z}^{(1)}_\lambda = \mathcal{Z}^{(2)}_\lambda =
    \mathcal{Z}_\lambda,
    \qquad
    \mathcal{Y}_\lambda = 
    \mathcal{Z} - \mathcal{Z}_\lambda + 2 \vartheta_{1,\lambda}, 
    \qquad
    \mathcal{X}_\lambda = 0.
  \end{array}
\end{equation}
The OPEs are then given by
\begin{eqnarray}
  \mu_\lambda(x)\mu_{1-\lambda} &=& 
  - \frac{\mathcal{O}_\lambda}{\mathcal{O}}
  |x|^{2\lambda(1-\lambda)} 
  \Bigl(
    \omega + (\ln|x|^2 + 2\mathcal{Z} - \mathcal{Z}_\lambda) \Omega
  \Bigr), 
  \\*
  \Lambda^{-2}
  \mu_\lambda(x)\Omega &=& 
  \mu_\lambda, 
  \\*
  \Lambda^{-2}
  \mu_\lambda(x)\omega &=& 
  - 
  \Bigl(
    \ln|x|^2 + \mathcal{Z}_\lambda
  \Bigr)
  \mu_\lambda. 
\end{eqnarray}

\subsection{Three twist fields}
\label{app:tloc3}

For the three-point amplitude
$\langle\mu_{\lambda_1}\mu_{\lambda_2}\mu_{\lambda_3}\rangle$ we have
to consider two cases. If $\lambda_1+\lambda_2+\lambda_3=1$ the
amplitude satisfies
\begin{equation}
  \left[
    \partial_1 
    -
    \lambda_1 \left(
      \frac{\lambda_2}{z_{12}} + 
      \frac{\lambda_3}{z_{13}} 
    \right)
  \right]
  \langle\mu_{\lambda_1}\mu_{\lambda_2}\mu_{\lambda_3}\rangle
  = 0.
\end{equation}
while for $\lambda_1+\lambda_2+\lambda_3=2$, we have
\begin{equation}
  \left[
    \partial_1 
    -
    (1-\lambda_1) \left(
      \frac{1-\lambda_2}{z_{12}} + 
      \frac{1-\lambda_3}{z_{13}} 
    \right)
  \right]
  \langle\mu_{\lambda_1}\mu_{\lambda_2}\mu_{\lambda_3}\rangle
  = 0.
\end{equation}
From this we obtain the amplitudes
\begin{equation}
  \renewcommand{\arraystretch}{1.5}
  \begin{array}{rcll}
    \langle\mu_{\lambda_1}\mu_{\lambda_2}\mu_{\lambda_3}\rangle 
    &=&
    \mathcal{C}_{\lambda_1,\lambda_2,\lambda_3}
    \left|
      z_{12}^{\lambda_1\lambda_2}
      z_{13}^{\lambda_1\lambda_3}
      z_{23}^{\lambda_2\lambda_3}
    \right|^2
    &
    \hbox{for $\lambda_1+\lambda_2+\lambda_3=1$} \,, 
    \\
    &=&
    \mathcal{C}_{\lambda_1,\lambda_2,\lambda_3}
    \left|
      z_{12}^{\lambda_1^*\lambda_2^*}
      z_{13}^{\lambda_1^*\lambda_3^*}
      z_{23}^{\lambda_2^*\lambda_3^*}
    \right|^2
    &
    \hbox{for $\lambda_1+\lambda_2+\lambda_3=2$} \,, 
  \end{array}
\end{equation}

\subsection{Four twist fields}
\label{app:tloc4}

In the case of four twist fields we obtain
\begin{equation}
  \left[
    \partial_1 
    -
    \lambda_1 \left(
      \frac{\lambda_2}{z_{12}} + 
      \frac{\lambda_3}{z_{13}} + 
      \frac{\lambda_4}{z_{14}} 
    \right)
  \right]
  \langle\mu_{\lambda_1}\mu_{\lambda_2}\mu_{\lambda_3}\mu_{\lambda_4}\rangle
  = 0.
\end{equation}
for $\lambda_1+\lambda_2+\lambda_3+\lambda_4=1$ and 
\begin{equation}
  \left[
    \partial_1 
    -
    (1-\lambda_1) \left(
      \frac{1-\lambda_2}{z_{12}} + 
      \frac{1-\lambda_3}{z_{13}} + 
      \frac{1-\lambda_4}{z_{14}} 
    \right)
  \right]
  \langle\mu_{\lambda_1}\mu_{\lambda_2}\mu_{\lambda_3}\mu_{\lambda_4}\rangle
  = 0.
\end{equation}
for $\lambda_1+\lambda_2+\lambda_3+\lambda_4=3$.
The four-point amplitudes, manifestly symmetric in the fields, are thus
\begin{equation}
  \begin{array}{rcll}
    \langle\mu_{\lambda_1}\mu_{\lambda_2}\mu_{\lambda_3}
    \mu_{\lambda_4}\rangle  
    &=&
    \mathcal{F}_{\lambda_1,\lambda_2,\lambda_3,\lambda_4}
    \left|
      \prod_{i<j} z_{ij}^{\lambda_i\lambda_j}
    \right|^2
    &
    \hbox{for $\lambda_1+\lambda_2+\lambda_3+\lambda_4=1$}
    \,, 
    \\
    &=&
    \mathcal{F}_{\lambda_1,\lambda_2,\lambda_3,\lambda_4}
    \left|
      \prod_{i<j} z_{ij}^{\lambda_i^*\lambda_j^*}
    \right|^2
    &
    \hbox{for $\lambda_1+\lambda_2+\lambda_3+\lambda_4=3$}
    \,.
  \end{array}
\end{equation}
The last case, $\lambda_1+\lambda_2+\lambda_3+\lambda_4=2$, requires
a second order differential equation. Using
$\alpha_1=2-\lambda_1, \alpha_2=-\lambda_2, \alpha_3=-\lambda_3,
\alpha_4=-\lambda_4$ and $\alpha_1=1+\lambda_1, \alpha_2=\lambda_2-1,
\alpha_3=\lambda_3-1, \alpha_4=\lambda_4-1$ we obtain
\begin{eqnarray*}
  &&
  \Biggl\{
  \partial_1^2 
  +
  \left[
    \lambda_1^* \left(
      \frac{\lambda_2}{z_{12}} + \frac{\lambda_3}{z_{13}} +
      \frac{\lambda_4}{z_{14}} 
    \right)
    +
    \lambda_1 \left(
      \frac{\lambda_2^*}{z_{12}} + \frac{\lambda_3^*}{z_{13}} +
      \frac{\lambda_4^*}{z_{14}} 
    \right)
  \right] \partial_1
  \\*&&
  +
    \lambda_1\lambda_1^* \left(
      \frac{\lambda_2\lambda_2^*}{z_{12}^2} + 
      \frac{\lambda_3\lambda_3^*}{z_{13}^2} +
      \frac{\lambda_4\lambda_4^*}{z_{14}^2} 
      -
      \frac{\lambda_2\lambda_3+\lambda_2^*\lambda_3^*}{z_{12}z_{13}} - 
      \frac{\lambda_2\lambda_4+\lambda_2^*\lambda_4^*}{z_{14}z_{12}} -
      \frac{\lambda_3\lambda_4+\lambda_3^*\lambda_4^*}{z_{13}z_{14}} 
    \right)
  \Biggr\}
  \\*&&\times
  \langle\mu_{\lambda_1}\mu_{\lambda_2}\mu_{\lambda_3}\mu_{\lambda_4}\rangle
  = 0
\end{eqnarray*}
M\"obius covariance implies the amplitude can be written as
\begin{eqnarray}
  \lefteqn{
  \langle\mu_{\lambda_1}\mu_{\lambda_2}\mu_{\lambda_3}\mu_{\lambda_4}\rangle
    = } \qquad \nonumber\\*
  &=&
  \left|
    z_{12}^{-h_1-h_2+h_3+h_4} z_{13}^{-2h_3} 
    z_{14}^{-h_1+h_2+h_3-h_4} z_{24}^{h_1-h_2-h_3-h_4}
  \right|^2
  \left|
    x^{\lambda_3\lambda_4} (1-x)^{\lambda_2\lambda_3}
  \right|^2 f(x,\bar x),
  \nonumber\\*
  &=&
  \left|
    z_{12}^{\lambda_1^*\lambda_2^*}
    z_{34}^{\lambda_3\lambda_4}
    z_{14}^{\lambda_1^*\lambda_4^*}
    z_{23}^{\lambda_2\lambda_3}
    z_{13}^{\lambda_1^*\lambda_3^*}
    z_{24}^{\lambda_2\lambda_4}
    (z_{13}z_{24})^{-\lambda_1^*}
  \right|^2 f(x,\bar x),
\end{eqnarray}
where $f(x,\bar x)$ satisfies the hypergeometric equation of type
$(\lambda_3, 1-\lambda_1, \lambda_3+\lambda_4)$, 
\begin{displaymath}
  x(1-x) f'' + \left[
    \lambda_3+\lambda_4 - \left(
      \lambda_2+2\lambda_3+\lambda_4 
    \right) x
  \right] f' - 
  \lambda_3 \left(
    \lambda_2+\lambda_3+\lambda_4-1
  \right) f = 0.
\end{displaymath}
If no two $\lambda_j$ add up to an integer, a system of two linear independent
solutions in the vicinity of $x=0$ is given by
\begin{equation}
  \label{eq:fi0}
  \renewcommand{\arraystretch}{1.5}
  \begin{array}{rcl}
    f_1^{(0)} &=& 
    {}_2F_1(\lambda_3, 1-\lambda_1; \lambda_3+\lambda_4; x)
    \,, \\*
    f_2^{(0)} &=& 
    x^{1-\lambda_3-\lambda_4} (1-x)^{1-\lambda_2-\lambda_3}
    {}_2F_1(\lambda_1, 1-\lambda_3; \lambda_1+\lambda_2; x)
    \,.
  \end{array}
\end{equation}
Another set of solutions, adapted to $x\sim1$, is 
\begin{equation}
  \label{eq:fi1}
  \renewcommand{\arraystretch}{1.5}
  \begin{array}{rcl}
    f_1^{(1)} &=& 
    {}_2F_1(\lambda_3, 1-\lambda_1; \lambda_2+\lambda_3; 1-x)
    \,, \\*
    f_2^{(1)} &=& 
    x^{1-\lambda_3-\lambda_4} (1-x)^{1-\lambda_2-\lambda_3}
    {}_2F_1(\lambda_1, 1-\lambda_3; \lambda_1+\lambda_4; 1-x)
    \,,
  \end{array}
\end{equation}
while a set of solutions in the vicinity of $x=\infty$ is given by
\begin{equation}
  \label{eq:fii}
  \renewcommand{\arraystretch}{1.5}
  \begin{array}{rcl}
    f_1^{(\infty)} &=& 
    x^{-\lambda_3} 
    {}_2F_1(\lambda_3, 1-\lambda_4; \lambda_1+\lambda_3; 1/x)
    \,, \\*
    f_2^{(\infty)} &=& 
    x^{-\lambda_4} (1-x)^{1-\lambda_2-\lambda_3}
    {}_2F_1(\lambda_4, 1-\lambda_3; \lambda_2+\lambda_4; 1/x)
    \,.
  \end{array}
\end{equation}
Monodromy invariance requires that the non-chiral solution is of the
form, up to an overall constant, 
\begin{eqnarray*}
  f(x,\bar x) 
  &=& 
  - \rho(\lambda_3,\lambda_4) |f_1^{(0)}|^2 
  - \rho(\lambda_1,\lambda_2) |f_2^{(0)}|^2 
  \\*
  &=& 
  - \rho(\lambda_2,\lambda_3) |f_1^{(1)}|^2 
  - \rho(\lambda_1,\lambda_4) |f_2^{(1)}|^2
  \\*
  &=& 
  - \rho(\lambda_1,\lambda_3) |f_1^{(\infty)}|^2 
  - \rho(\lambda_2,\lambda_4) |f_2^{(\infty)}|^2 
  \,,
\end{eqnarray*}
where
\begin{displaymath}
  \rho(\lambda,\lambda') = 
  \frac{\Gamma(1-\lambda-\lambda')}{\Gamma(\lambda+\lambda')}
  \frac{\Gamma(\lambda)}{\Gamma(1-\lambda)}
  \frac{\Gamma(\lambda')}{\Gamma(1-\lambda')}
  =
  \rho(\lambda,1-\lambda-\lambda')
  \,.
\end{displaymath}
Note that, if $\lambda_1+\lambda_2+\lambda_3+\lambda_4=2$ we have
\begin{displaymath}
  \frac{\rho(1-\lambda_1,1-\lambda_2)}{\rho(\lambda_3,\lambda_4)} = 
  \frac{\rho(1-\lambda_3,1-\lambda_4)}{\rho(\lambda_1,\lambda_2)} = 
  \frac{\Gamma(\lambda_1)\Gamma(\lambda_2)
    \Gamma(\lambda_3)\Gamma(\lambda_4)} 
  {\Gamma(1-\lambda_1)\Gamma(1-\lambda_2)
  \Gamma(1-\lambda_3)\Gamma(1-\lambda_4)} \,.
\end{displaymath}
Thus the full
four-point amplitude, manifestly symmetric in the fields, can be
written as 
\begin{eqnarray}
  \lefteqn{
    \mathcal{F}_{\lambda_1,\lambda_2,\lambda_3,\lambda_4}^{-1}
    \langle\mu_{\lambda_1}\mu_{\lambda_2}\mu_{\lambda_3}\mu_{\lambda_4}\rangle
    = } \qquad \nonumber\\*
  &=&
  - 
  \sqrt{\rho(\lambda_3,\lambda_4)\rho(\lambda_1^*,\lambda_2^*)}
  \left|
    z_{12}^{\lambda_1^*\lambda_2^*}
    z_{34}^{\lambda_3\lambda_4}
    z_{14}^{\lambda_1^*\lambda_4^*}
    z_{23}^{\lambda_2\lambda_3}
    z_{13}^{\lambda_1^*\lambda_3^*}
    z_{24}^{\lambda_2\lambda_4}
    (z_{13}z_{24})^{-\lambda_1^*}
  \right|^2
  \nonumber\\*&&\quad\times
  \left|
    {}_2F_1(\lambda_3, 1-\lambda_1; \lambda_3+\lambda_4; x)
  \right|^2
  \nonumber\\*&&
  -
  \sqrt{\rho(\lambda_1,\lambda_2)\rho(\lambda_3^*,\lambda_4^*)}
  \left|
    z_{12}^{\lambda_1\lambda_2}
    z_{34}^{\lambda_3^*\lambda_4^*}
    z_{14}^{\lambda_1\lambda_4}
    z_{23}^{\lambda_2^*\lambda_3^*}
    z_{13}^{\lambda_1\lambda_3}
    z_{24}^{\lambda_2^*\lambda_4^*}
    (z_{13}z_{24})^{-\lambda_3^*}
  \right|^2
  \nonumber\\*&&\quad\times
  \left|
    {}_2F_1(\lambda_1, 1-\lambda_3; \lambda_1+\lambda_2; x)
  \right|^2
  \label{eq:mmmm2f0}
  \\
  &=& 
  - 
  \sqrt{\rho(\lambda_2,\lambda_3)\rho(\lambda_1^*,\lambda_4^*)}
  \left|
    z_{12}^{\lambda_1^*\lambda_2^*}
    z_{34}^{\lambda_3\lambda_4}
    z_{14}^{\lambda_1^*\lambda_4^*}
    z_{23}^{\lambda_2\lambda_3}
    z_{13}^{\lambda_1^*\lambda_3^*}
    z_{24}^{\lambda_2\lambda_4}
    (z_{13}z_{24})^{-\lambda_1^*}
  \right|^2
  \nonumber\\*&&\quad\times
  \left|
    {}_2F_1(\lambda_3, 1-\lambda_1; \lambda_2+\lambda_3; 1-x)
  \right|^2
  \nonumber\\*&&
  -
  \sqrt{\rho(\lambda_1,\lambda_4)\rho(\lambda_2^*,\lambda_3^*)}
  \left|
    z_{12}^{\lambda_1\lambda_2}
    z_{34}^{\lambda_3^*\lambda_4^*}
    z_{14}^{\lambda_1\lambda_4}
    z_{23}^{\lambda_2^*\lambda_3^*}
    z_{13}^{\lambda_1\lambda_3}
    z_{24}^{\lambda_2^*\lambda_4^*}
    (z_{13}z_{24})^{-\lambda_3^*}
  \right|^2
  \nonumber\\*&&\quad\times
  \left|
    {}_2F_1(\lambda_1, 1-\lambda_3; \lambda_1+\lambda_4; 1-x)
  \right|^2
  \label{eq:mmmm2f1}
  \\
  &=& 
  - 
  \sqrt{\rho(\lambda_1,\lambda_3)\rho(\lambda_1^*,\lambda_4^*)}
  \left|
    z_{12}^{\lambda_1\lambda_2}
    z_{34}^{\lambda_3^*\lambda_4^*}
    z_{14}^{\lambda_1^*\lambda_4^*}
    z_{23}^{\lambda_2\lambda_3}
    z_{13}^{\lambda_1\lambda_3}
    z_{24}^{\lambda_2^*\lambda_4^*}
    (z_{12}z_{34})^{-\lambda_4^*}
  \right|^2
  \nonumber\\*&&\quad\times
  \left|
    {}_2F_1(\lambda_3, 1-\lambda_4; \lambda_1+\lambda_3; 1/x)
  \right|^2
  \nonumber\\*&&
  -
  \sqrt{\rho(\lambda_1,\lambda_4)\rho(\lambda_1^*,\lambda_3^*)}
  \left|
    z_{12}^{\lambda_1\lambda_2}
    z_{34}^{\lambda_3^*\lambda_4^*}
    z_{14}^{\lambda_1\lambda_4}
    z_{23}^{\lambda_2^*\lambda_3^*}
    z_{13}^{\lambda_1^*\lambda_3^*}
    z_{24}^{\lambda_2\lambda_4}
    (z_{13}z_{24})^{-\lambda_3^*}
  \right|^2
  \nonumber\\*&&\quad\times
  \left|
    {}_2F_1(\lambda_4, 1-\lambda_3; \lambda_2+\lambda_4; 1/x)
  \right|^2
  \label{eq:mmmm2fi}
\end{eqnarray}
in the vicinity of $x\sim0, 1$ and $\infty$, respectively. The
constant $\mathcal{F}_{\lambda_1,\lambda_2,\lambda_3,\lambda_4}$ is
independent of the ordering of the indices.

If two $\lambda_j$ add up to an integer, $\lambda_1+\lambda_4=1,
\lambda_2+\lambda_3=1, \lambda_3+\lambda_4\neq1$ say, we encounter a
degenerate case of the hypergeometric equation. We still have the same
system of solutions near $x=0$,
\begin{eqnarray*}
  f_1^{(0)} &=& 
  {}_2F_1(\lambda_3, \lambda_4; \lambda_3+\lambda_4; x)
  \,, \\*
  f_2^{(0)} &=& 
  x^{1-\lambda_3-\lambda_4} 
  {}_2F_1(1-\lambda_3, 1-\lambda_4; 2-\lambda_3-\lambda_4; x)
  \,.
\end{eqnarray*}
Their analytic continuation to $x\sim1$ is given by
\begin{eqnarray*}
  f_1^{(0)} &=& 
  - \frac{\Gamma(\lambda_3+\lambda_4)}
  {\Gamma(\lambda_3)\Gamma(\lambda_4)} 
  \biggl(
    f^{(1)}
    [ \ln(1-x) - \vartheta_{\lambda_3,\lambda_4} ]
    + 
    \tilde f^{(1)}
  \biggr)
  \,, \\*
  f_2^{(0)} 
  &=& 
  - \frac
  {\Gamma(2-\lambda_3-\lambda_4)}
  {\Gamma(1-\lambda_3)\Gamma(1-\lambda_4)} 
  \biggl(
    f^{(1)}
    [ \ln(1-x) - \vartheta_{1-\lambda_3,1-\lambda_4} ]
    + 
    \tilde f^{(1)}
  \biggr)
  \,,
\end{eqnarray*}
where
\begin{eqnarray*}
  f_1 &=& {}_2F_1(\lambda_3, \lambda_4; 1; 1-x) 
  \,,
  \\*
  \tilde f_1 &=& M(\lambda_3, \lambda_4; 1; 1-x) 
  \,,
  \\
  \vartheta_{a,b} &=& 2 \psi(1) - \psi(a) - \psi(b) 
  \,, \\*
  M(a,b;c;x) &=& 
  \sum_{n=1}^\infty
  \frac{(a)_n (b)_n}{(c)_n n!}
  \left[
    h_{a,n} + h_{b,n} - h_{c,n} - h_{1,n}
  \right] x^n
  \,, \\*
  h_{a,n} &=& \psi(a+n) - \psi(a)
  \,,
\end{eqnarray*}
and $\psi(x)=\Gamma(x)'/\Gamma(x)$ is the digamma function.
The monodromy invariant solution is
\begin{eqnarray*}
  f(x,\bar x) 
  &=& -\rho(\lambda_3,\lambda_4) |f_1^{(0)}|^2 
  - \rho(\lambda_1,\lambda_2) |f_2^{(0)}|^2 
  \\*
  &=& 
  \left|
    {}_2F_1(\lambda_3, \lambda_4; 1; 1-x)
  \right|^2
  \left(
    \ln|1-x|^2
    - 
    \vartheta_{\lambda_3,1-\lambda_3}
    - 
    \vartheta_{\lambda_4,1-\lambda_4}
  \right)
  \\*&&{}
  +
  \left(
    {}_2F_1(\lambda_3, \lambda_4; 1; 1-x)
    M(\lambda_3, \lambda_4; 1; 1-\bar x)
    +
    \textrm{c.c.}
  \right)
  \,.
\end{eqnarray*}
If $\lambda_3\neq\lambda_4$, that is we have two different pairs of
conjugate fields, the solution near $x\sim\infty$ is still given by 
(\ref{eq:mmmm2fi}). If $\lambda_3=\lambda_4=\lambda$ then 
\begin{eqnarray*}
  f(x,\bar x) 
  &=& 
  |x|^{-2\lambda}
  \Biggl[
  \left|
    {}_2F_1(\lambda, 1-\lambda; 1; 1/x)
  \right|^2
  \left(
    - \ln|x|^2
    - 
    2 \vartheta_{\lambda,1-\lambda}
  \right)
  \\*&&\qquad{}
  +
  \left(
    {}_2F_1(\lambda, 1-\lambda; 1; 1/x)
    M(\lambda, 1-\lambda; 1; 1/\bar x)
    +
    \textrm{c.c.}
  \right)
  \Biggr]
  \,.
\end{eqnarray*}
One can deal similarly with the other cases where we have one or more
pairs of conjugate fields. The generic solution for all cases is given
by (\ref{eq:mmmm2f0}) -- (\ref{eq:mmmm2fi}), which are replaced by a
degenerate solution in the following cases, respectively:
\begin{description}
\item[$\lambda_1+\lambda_2=\lambda_3+\lambda_4=1$:] 
  \begin{eqnarray*}
    \label{eq:mmmm2d0}
    \langle\mu_{\lambda_1}\mu_{\lambda_2}\mu_{\lambda_3}\mu_{\lambda_4}\rangle
    &=& 
    \mathcal{F}_{\lambda_1,\lambda_2,\lambda_3,\lambda_4}
    |z_{12}|^{2\lambda_1\lambda_2}
    |z_{34}|^{2\lambda_3\lambda_4}
    \left|\frac{z_{14}z_{23}}{z_{13}z_{24}}\right|^{2\lambda_2\lambda_3} 
    \\*&&\times
    \biggl[
    \left|
      {}_2F_1(\lambda_2, \lambda_3; 1; x)
    \right|^2
    \left(
      \ln|x|^2
      - 
      \vartheta_{\lambda_1,\lambda_2}
      - 
      \vartheta_{\lambda_3,\lambda_4}
    \right)
    \\*&&{}
    +
    \left(
      {}_2F_1(\lambda_2, \lambda_3; 1; x)
      M(\lambda_2, \lambda_3; 1; \bar x)
      +
      \textrm{c.c.}
    \right)
    \biggr]
  \end{eqnarray*}
\item[$\lambda_1+\lambda_4=\lambda_2+\lambda_3=1$:] 
  \begin{eqnarray*}
    \label{eq:mmmm2d1}
    \langle\mu_{\lambda_1}\mu_{\lambda_2}\mu_{\lambda_3}\mu_{\lambda_4}\rangle
    &=& 
    \mathcal{F}_{\lambda_1,\lambda_2,\lambda_3,\lambda_4}
    |z_{14}|^{2\lambda_1\lambda_4}
    |z_{23}|^{2\lambda_2\lambda_3}
    \left|\frac{z_{12}z_{34}}{z_{13}z_{24}}\right|^{2\lambda_3\lambda_4} 
    \\*&&\times
    \biggl[
    \left|
      {}_2F_1(\lambda_3, \lambda_4; 1; 1-x)
    \right|^2
    \left(
      \ln|1-x|^2
      - 
      \vartheta_{\lambda_1,\lambda_4}
      - 
      \vartheta_{\lambda_2,\lambda_3}
    \right)
    \\*&&\quad{}
    +
    \left(
      {}_2F_1(\lambda_3, \lambda_4; 1; 1-x)
      M(\lambda_3, \lambda_4; 1; 1-\bar x)
      +
      \textrm{c.c.}
    \right)
    \biggr]
  \end{eqnarray*}
\item[$\lambda_1+\lambda_3=\lambda_2+\lambda_4=1$:] 
  \begin{eqnarray*}
    \label{eq:mmmm2di}
    \langle\mu_{\lambda_1}\mu_{\lambda_2}\mu_{\lambda_3}\mu_{\lambda_4}\rangle
    &=& 
    \mathcal{F}_{\lambda_1,\lambda_2,\lambda_3,\lambda_4}
    |z_{13}|^{2\lambda_1\lambda_3}
    |z_{24}|^{2\lambda_2\lambda_4}
    \left|\frac{z_{14}z_{23}}{z_{12}z_{34}}\right|^{2\lambda_2\lambda_3} 
    \\*&&\times
    \biggl[
    \left|
      {}_2F_1(\lambda_2, \lambda_3; 1; 1/x)
    \right|^2
    \left(
      - \ln|x|^2
      - 
      \vartheta_{\lambda_1,\lambda_3}
      - 
      \vartheta_{\lambda_2,\lambda_4}
    \right)
    \\*&&{}
    +
    \left(
      {}_2F_1(\lambda_2, \lambda_3; 1; 1/x)
      M(\lambda_2, \lambda_3; 1; 1/\bar x)
      +
      \textrm{c.c.}
    \right)
    \biggr]
  \end{eqnarray*}
\end{description}

\section{Character formulae}
\label{app:chars}

We derive here the characters of $\mathcal{A}_\mu$ in the
$\mathcal{C}_{2N}$ orbifold model. Realising $\mathcal{C}_{2N}$ as the
$2N$-th roots of unity
we can use the Jacobi triple product identity to obtain
\begin{eqnarray*}
  \chi_{\mathcal{A}}(\tau,u) &=& 
  q^{\frac{1}{12}} \prod_{n=1}^\infty (1+u q^n) (1+u^{-1}q^n)
  \\
  &=&
  \left\{
    {\renewcommand\arraystretch{2}
      \begin{array}{ll}
        \displaystyle\sum_{m=0}^{2N-1} 
        \frac{u^m}{1+u^{-1}} \Lambda_{(2m+1)N,2N^2}(\tau) &
        \textrm{for $u\neq-1$} \\
        \displaystyle\eta(\tau)^2 & \textrm{for $u=-1$} 
      \end{array}}
  \right.
  \\
  &=&
  \sum_{k=0}^{2N-1} u^k \chi_{\mathcal{A}_{k/2N}}(\tau),
\end{eqnarray*}
where $u\in\mathbb{C}$ with $u^{2N}=1$. The Virasoro characters of the
spaces $\mathcal{A}_{k/2N}$ can then be
obtained by
\begin{eqnarray*}
  \chi_{\mathcal{A}_{k/2N}}(\tau) 
  &=& 
  \frac{1}{2N} \sum_{l=0}^{2N-1} \omega^{-kl}
  \chi_{\mathcal{A}}(\tau,\omega^l) 
  \\
  &=&
  \frac{1}{2N} \left[
    (-1)^k \eta(\tau)^2 + \sum_{m=0}^{2N-1} \left(
      \sum_{l=-N+1}^{N-1} \frac{\omega^{(m-k)l}}{1+\omega^{-l}}
    \right) \Lambda_{(2m+1)N,2N^2}(\tau) 
  \right]
  \\
  &=&
  \frac{1}{2N} \left[
    (-1)^k \eta(\tau)^2 + \sum_{m=0}^{N-1} \left(
      \sum_{l=-N+1}^{N-1} \omega^{-kl}
      \frac{\omega^{ml}+\omega^{-(m+1)l}}{1+\omega^{-l}}
    \right) \Lambda_{(2m+1)N,2N^2}(\tau) 
  \right]
\end{eqnarray*}
where $\omega=\exp(\pi\i/N)$ is the primitive $2N$-th root of
unity. The coefficients can be evaluated further as
\begin{eqnarray*}
  \sum_{l=-N+1}^{N-1} \omega^{-kl}
  \frac{\omega^{ml}+\omega^{-(m+1)l}}{1+\omega^{-l}}
  &=&
  \sum_{l=-N+1}^{N-1} \omega^{-kl}
  \sum_{r=-m}^m (-1)^{r+m} \omega^{rl}
  \\
  &=&
  \sum_{r=-m}^m (-1)^{r+m} \sum_{l=-N+1}^{N-1} \omega^{(r-k)l}
  \\
  &=&
  \sum_{r=-m}^m (-1)^{r+m} \left(
    2N \delta_{r,k} - (-1)^{r-k} 
  \right)
  \\
  &=&
  \left\{
    {\renewcommand\arraystretch{2}
      \begin{array}{ll}
        \displaystyle 
        (-1)^{m-k} (2N-2m-1) & 
        \textrm{for $|k|\leq m$} \\
        \displaystyle 
        (-1)^{m-k+1} (2m+1) & 
        \textrm{for $|k|>m$} 
      \end{array}}
  \right.,
\end{eqnarray*}
where we assumed $|k|\leq N$.
Thus the characters can be written as
\begin{eqnarray*}
  \chi_{\mathcal{A}_{k/2N}}(\tau) 
  &=& 
  \frac{(-1)^k}{2N} \Biggl[
    \eta(\tau)^2 + 
    \sum_{l=0}^{|k|-1} (-1)^{l+1} (2l+1)
    \Lambda_{(2l+1)N,2N^2}(\tau) 
    \\*&&\qquad{}+ 
    \sum_{l=|k|}^{N-1} (-1)^{l} (2N-2l-1)
    \Lambda_{(2l+1)N,2N^2}(\tau) 
  \Biggr]
\end{eqnarray*}
                                

%
\end{document}